\documentclass[a4paper,fleqn,usenatbib]{mnras}
\usepackage[utf8]{inputenc}
\usepackage[catalan, english]{babel}
\usepackage{geometry}
\usepackage{graphicx}
\graphicspath{{./figures/}}
\usepackage{natbib}
\usepackage{appendix}
\usepackage{tabularx}
\usepackage[catalan,english]{varioref}
\vrefwarning
\usepackage{color}
\usepackage{booktabs}
\usepackage{threeparttable}
\usepackage{longtable}
\usepackage{multicol}
\usepackage{multirow}
\usepackage{enumerate}
\usepackage{amsmath}
\usepackage{amssymb}
\usepackage{setspace} 
\usepackage{hyperref}
\usepackage[english]{cleveref}

\definecolor{halfgray}{gray}{0.55}
\definecolor{naviblue}{RGB}{0,0,102}
\definecolor{webbrown}{rgb}{.6,0,0}
\definecolor{RoyalBlue}{cmyk}{1, 0.50, 0, 0}
\definecolor{webgreen}{rgb}{0,.5,0}
\definecolor{Maroon}{cmyk}{0, 0.87, 0.68, 0.32}
\definecolor{Black}{cmyk}{0, 0, 0, 0}
\definecolor{myorange}{RGB}{239, 186,67}

\def\aap{A\&A}

\def\apjs{ApJS}
\def\apjl{ApJL}

\newcommand{\lya}{Ly$\alpha$}
\newcommand{\lyb}{Ly$\beta$}
\newcommand{\cm}{\,\mathrm{cm}}
\newcommand{\angs}{\textup{\AA}}

\newcommand{\kms}{\, {\rm km}\, {\rm s}^{-1} }

\newcommand{\hmpc}{\, h^{-1} {\rm Mpc}}
\newcommand{\hmsun}{\, h^{-1} {\rm M_{\odot}}}

\newcommand{\ovi}{\rm O\thinspace VI}

\crefname{equation}{equation}{equations}
\crefname{figure}{figure}{figures}
\crefname{appsec}{appendix}{appendices}
\creflabelformat{equation}{#2#1#3}

\begin{document}

\title[DLA-\lya{} forest cross-correlation]{The SDSS-DR12 large-scale cross-correlation of Damped Lyman Alpha Systems with the Lyman Alpha Forest}
\date{}
\author[Ignasi P\'erez-R\`afols et al.]
  {Ignasi ~P\'erez-R\`afols,$^{1,2}$\thanks{email: iprafols@icc.ub.edu}, Andreu ~Font-Ribera$^{3}$, Jordi ~Miralda-Escud\'e$^{1, 4}$,
\newauthor
Michael ~Blomqvist$^{5}$, Simeon ~Bird$^{6}$, Nicol\'as ~Busca$^{7}$,
\newauthor
H\'elion ~du Mas des Bourboux$^{8}$, Llu\'is ~Mas-Ribas$^{9}$, Pasquier ~Noterdaeme$^{10}$,
\newauthor 
Patrick ~Petitjean$^{10}$, James ~Rich$^{8}$, Donald P. ~Schneider$^{11,12}$
\\
$^{1}$Institut de Ci\`encies del Cosmos, Universitat de Barcelona/IEEC, Barcelona, E-08028, Catalonia\\
$^{2}$Departament de F\'isica Qu\`antica i Astrof\'isica, Universitat de Barcelona/IEEC, Barcelona, E-08028, Catalonia\\
$^{3}$Department of Physics and Astronomy, University College London, Gower Street, London, United Kingdom\\
$^{4}$Instituci\'o Catalana de Recerca i Estudis Avan\c{c}ats, Barcelona, Catalonia\\
$^{5}$Aix Marseille Univ, CNRS, LAM, Laboratoire d'Astrophysique de Marseille, Marseille, France
\\
$^{6}$ Department of Physics \& Astronomy, Johns Hopkins University, 3400 N.~Charles Street, Baltimore, MD 21218, USA\\
$^{7}$APC, Universit\'e Paris Diderot-Paris 7, CNRS/IN2P3, CEA, Observatoire de Paris, 10, rue A. Domon \& L. Duquet, Paris, France\\
$^{8}$IRFU, CEA, Universit\'e Paris-Saclay,  F-91191 Gif-sur-Yvette, France\\
$^{9}$Institute of Theoretical Astrophysics, University of Oslo, P.O. Box 1029 Blindern, NO-0315    Oslo, Norway\\
$^{10}$Universite\'e Paris 6 et CNRS, Institut d’Astrophysique de Paris, 98bis blvd. Arago, 75014 Paris, France\\
$^{11}$Department of Astronomy and Astrophysics, The Pennsylvania State University, University Park, PA 16802\\
$^{12}$Institute for Gravitation and the Cosmos, The Pennsylvania State University, University Park, PA 16802
}

\maketitle

\begin{abstract}
  
 We present a measurement of the DLA mean bias from the
cross-correlation of DLA and the \lya{} forest, updating earlier
results of \cite{Font-Ribera+2012} with the final BOSS Data Release and
an improved method to address continuum fitting corrections. Our
cross-correlation is well fitted by linear theory with the standard
$\Lambda CDM$ model, with a DLA bias of $b_{\rm DLA} = 1.99\pm 0.11$; a
more conservative analysis, which removes DLA in the \lyb{} forest and
uses only the cross-correlation at $r> 10\hmpc$, yields
$b_{\rm DLA} = 2.00\pm 0.19$. This assumes the cosmological model from
\cite{Planck2015} and the \lya{} forest bias factors of \cite{Bautista+2017},
and includes only statistical errors obtained from bootstrap analysis. The main
systematic errors arise from possible impurities and selection effects in the
DLA catalogue, and from uncertainties in the determination of the \lya{} forest
bias factors and a correction for effects of high column density absorbers. We
find no dependence of the DLA bias on column density or redshift. The
measured bias value corresponds to a host halo mass $\sim 4\cdot10^{11} \hmsun$
if all DLA were hosted in halos of a similar mass. In a realistic
model where host halos over a broad mass range have a DLA cross section
$\Sigma(M_h) \propto M_h^{\alpha}$ down to $M_h > M_{\rm min} =10^{8.5} \hmsun$,
we find that $\alpha > 1$ is required to have $b_{\rm DLA}> 1.7$, implying a
steeper relation or higher value of $M_{\rm min}$ than is generally predicted
in numerical simulations of galaxy formation.

\end{abstract}
 

%
\section{Introduction}  \label{sec DLA: Introduction}

  Damped \lya{} Absorbers (DLAs) are absorption systems of high neutral hydrogen column
density, usually defined as $N_{HI}\ge 2\times10^{20}\cm^{-2}$
\cite{Wolfe+1986}. At these column densities, the damped profile of
the hydrogen \lya{} line is measurable even in low resolution spectra
and with the superposition of the \lya\ forest, allowing the column
density to be directly measured from the absorption profile.
This lower limit on $N_{HI}$ is also related (depending
on the ionization parameter, or ratio of the gas density to the
photoionization rate) to absorption systems which, owing to
self-shielding of the external cosmic ionizing background radiation,
have most of their hydrogen in atomic form
\citep[e.g.,][]{Vladilo+2001}. For reviews on DLAs
see, e.g., \cite{Wolfe+2005, Barnes+2014}.

  DLAs are therefore a probe to any gaseous systems that have
condensed to high enough densities to become self-shielding, which are
naturally associated with sites of galaxy formation. 
In the standard Cold Dark Matter (CDM) model of structure formation, we
expect these sites to be located in halos over a broad range of mass,
from those of dwarf galaxies to groups of massive galaxies. Measurements
of the incidence rate and column density distribution imply a
contribution to the matter density of the atomic gas contained in these
systems of $\Omega_{\rm DLA}\simeq10^{-3}$ at redshifts $2<z<3.5$
\citep{Peroux+2003,Prochaska+2005,Zafar+2013,Crighton+2015,
Padmanabhan+2016,Prochaska+2009,Noterdaeme+2009,Noterdaeme+2012b}. This
accounts for $\sim$ 2\% of all baryons in the universe, which is
comparable to the fraction of baryons in stars at the same redshifts.
These absorption systems are therefore regarded as reservoirs of atomic
gas clouds for the formation of the stellar component of galaxies, and
they are crucial to understand how galaxies can be gradually formed from
gas that is accreted in galactic halos.

  The study of metal absorption lines associated with DLAs is a
powerful tool to study the dynamics and evolution of this gas, and has revealed
that DLAs typically have low metallicities distributed over a
broad range of $10^{-3}Z_{\odot}$ to $1Z_{\odot}$, and on average decreasing
gradually with redshift
\citep{Kulkarni+2002, Vladillo2002,Prochaska+2003a, Kulkarni+2005, Rafelski+2012, Jorgenson+2013, Neeleman+2013, Moller+2013, Mas-Ribas+2017}.This implies that the gas reservoir in DLAs has been enriched
from material ejected by stars, which were formed either in low-mass
galaxies that later merged into the DLA host halo together with the gas,
or in a galaxy in the DLA host halo itself.
Absorption lines from low and high-ionization species associated with
DLAs suggest a broad range of densities and temperatures
\citep{Wolfe+2000, Prochaska+2002, Fox+2007a, Fox+2007b}. The
kinematics of these low and high ionization gas phases differ, and a
complex structure of absorption components at different velocities are
often seen in high spectral resolution data, reflecting a clumpy
structure with typical velocity ranges of $\sim 100\kms$
\citep{Prochaska+1997, Prochaska+1998, Wolfe+1998}.
Several models of gaseous galactic halos have been proposed to account
for these observations
\citep[see e.g.][]{Haehnelt+1998, McDonald+1999, Fumagalli+2011, Cen2012, Rahmati+2014, Bird+2015, Neeleman+2015}.

  Despite this rich information on the velocity structure of DLAs, the mass distribution of their host halo masses is not well-known. One way to characterize this distribution is to analyse the clustering properties of DLAs. In the limit of
large scales, where linear theory holds, the correlation function of any
population of objects that trace the primordial mass perturbations is
equal to the mass autocorrelation times the square of the bias factor
\citep[e.g.,][]{Cole+1989,Mo+1996b}.
In redshift space, where all our observations
are done, the same relation holds adding a redshift space distortion
term \citep{Kaiser1987}. The bias factor of halos increases with their
mass in a way that can be accurately predicted both analytically
\cite[see e.g.][]{Sheth+1999} and from sophisticated numerical
simulations \citep[see e.g.,][]{Tinker+2010}. Therefore, if every DLA
is associated with a dark matter halo, a measurement
of the mean bias factor of any population of DLAs tells us the mean
bias factor of their host halos and constrains in a powerful way their
mass distribution.

  A first method for measuring the DLA bias, $b_{\rm DLA}$, is by
measuring the DLA autocorrelation. This approach, however,
requires a large sample and has not been attempted so far due to smaller
number of DLAs compared to quasars. A more
convenient method is to use the cross-correlation with another tracer
population. The first
cross-correlation that was detected was with Lyman break galaxies in the
vicinity of the quasar lines of sight \citep{Cooke+2006}, but owing to
their small sample size (only 11 DLAs), the bias factor could
only be constrained to $1.3<b_{\rm DLA}<4$.

  The Baryon Oscillations Spectroscopic Survey (BOSS)  \citep{Dawson+2013} in the Sloan Digital Sky Survey III (SDSS-III),
\cite{Eisenstein+2011} allowed for a very large sample of quasars
and DLAs to be obtained, which opened the way for measuring a
variety of cross-correlations on scales much larger than had been
attainable before. The other tracer of cosmological density fluctuations
that is most useful for obtaining the DLA bias factor turns out to
be the \lya\ forest absorption, because of its presence in every
quasar spectrum over a broad redshift range. The cross-correlation with
the \lya{} forest was first measured by \citet[][hereafter FR12]{Font-Ribera+2012} using the ninth Data Release (DR9) of BOSS, with a sample of 7,458 DLAs, and
a value $b_{\rm DLA}=2.17 \pm0.20$ was obtained, where the error reflects
only uncertainties from the observational determination of the
cross-correlation, and not from the model used to derive the bias. The
main modelling uncertainty lies in the bias and redshift distortion
parameter of the \lya\ forest, because only the product of bias factors of
the two tracer populations can be determined. In that work, the
first determination of the \lya\ forest bias factors by \cite{Slosar+2011}
was used. This measurement was based on the early data release of the
BOSS sample of quasar spectra containing the \lya\ forest.

  This paper is an update to the measurement of the cross-correlation of
DLAs and the \lya\ forest by FR12. Using the
entire DR12 sample, we can decrease the errorbars of this measurement
and we can better explore the dependence of the bias factor on the
DLA column density and the redshift evolution. A dependence of the
mean bias factor on any DLA properties can provide powerful
constraints on galaxy formation models and tests on the predictive
accuracy of cosmological numerical simulations \citep[e.g.][]{Bird+2014}. In addition, we use the improved estimate of the \lya\ forest
bias factors by \cite{Bautista+2017}, implying a substantial reduction
of our systematic errors in deriving the DLA bias as well.

  We start by describing the datasets used to derive the DLA bias
in section \ref{sec DLA: Sample data}. An improved estimator for the
cross-correlation is described in section \ref{sec DLA: Cross-correlation}.
Section \ref{sec DLA: Model} explains the model used to fit the DLA
bias. Then, in section \ref{sec DLA: Results} we present our results. A
detailed comparison with previous measurements and a study of the model dependencies of the DLA bias measurement is made in section \ref{sec DLA: comparison DR9}.
Finally, the cosmological implications for the halo masses hosting
DLAs are discussed in section \ref{sec DLA: Discussion}, and we summarize our conclusions
in section \ref{sec DLA: Conclusions}. Throughout this paper we use a flat $\Lambda$CDM cosmology, with
$\Omega_{m}=0.3156$, $\Omega_{b}=0.0492$, $h=0.6727$, $n_{s}=0.9645$,
and $\sigma_{8}=0.831$, as reported by \cite{Planck2015}.

%
\section{Data Sample}  \label{sec DLA: Sample data}

 In this section we describe the datasets used in this study, based on
the DR12 of SDSS-III
\citep{Gunn+1998,York+2000,Gunn+2006,Eisenstein+2011,Bolton+2012,Smee+2013},
which is the final Data Release of BOSS \citep{Dawson+2013}. 
The quasar target selection used in BOSS is summarized in
\cite{Ross+2012}, and combines different targeting methods described in
\cite{Yeche+2010,Kirkpatrick+2011,Bovy+2011}.

  We measure the cross-correlation of two tracers of the underlying
density field: the number density of DLAs and the \lya{}
absorption along a set of lines of sight. The DLAs used as
tracers are designated here as {\it DLA sample} and the quasar
lines of sight where the \lya{} absorption is measured are designated
as {\it \lya{} sample}. All the quasars used to find the DLAs
and measure the \lya\ absorption spectra are in the DR12Q catalogue
\citep{Paris+2017}.

\subsection{DLA Sample} \label{subs DLA: DLA Sample}

 For the DLA sample we use an early version of the DR12 extension
of the DLA catalogue from \cite{Noterdaeme+2012b}. This sample
contains a total of 34,050 DLAs candidates with column density
$N_{HI}\ge10^{20}\cm^{-2}$. For convenience, from here on we will refer to these DLAs candidates simply as DLAs. We note that the precise number of DLAs
varies slightly with the different versions of the catalogue that were
produced, but the inclusion or exclusion of the small number of objects
that differ among the versions does not affect in any significant way
the results in this paper. Although the strict definition of a
DLA requires its column density to be above
$2\times10^{20}\cm^{-2}$, systems with column density down to
$10^{20}\cm^{-2}$ are still identified with high efficiency in
BOSS data and are not expected to sharply change their nature.
We will test the dependence of the properties of DLAs we measure
with column density. Out of the 34,050 DLAs, there are 12 which
have the catalogue identifier {\it ThingID} set to $-1$, which indicates
an error in the pipeline data reduction for these objects. They are
excluded from the final sample. 

  We now describe several cuts we apply to the remaining 34,038
DLAs to obtain our DLA sample with an increased purity
compared to that of the catalogue. Purity of our sample is
important because objects included in the catalogue that are not real
or are at the wrong redshift will decrease the measured bias of
DLAs, while confusion with other types of absorption systems
(e.g., Lyman limit systems with extra \lya{} forest absorption around
them in high noise spectra) might increase the measured bias if these
absorption systems have a higher bias than DLAs. On the other
hand, completeness is less important: eliminating a fraction of the
real DLAs will only result in an increase of the errors of the
cross-correlation without modifying it systematically, as long as the
probability of inclusion of the DLAs in the catalogue is not
correlated with its large-scale cosmological environment. The cuts
applied here are the same as those in FR12, except
that we add additional ones to obtain different samples and test the
dependence of our results on them, and they are as follows:

 {\it First cut: DLA redshift, $z_{\rm DLA}$}. We include only DLAs in
the redshift range $2.0\le z_{\rm DLA}<3.5$. Outside this redshift interval,
DLAs have few nearby lines of sight with sufficient
signal-to-noise ratio in the \lya{} forest to be useful to measure the
correlation, and we eliminate them to have a well-defined redshift
interval. This reduces our sample to 31,059 DLAs.

  {\it Second cut: continuum-to-noise ratio (CNR) $\ge 3$}. The CNR
of the \lya{} forest spectral region, defined in \cite{Noterdaeme+2012b},
provides a good estimate of the data quality over the region of
interest, and is independent of the presence of DLAs. Since it is more difficult to detect DLAs in noisier spectra, we apply this
second cut to increase the purity of the sample without drastically
reducing the number of systems. A total of 23,568 DLAs survive
this cut. 

  {\it Third cut: eliminating Broad Absorption Line (BAL) systems}, which can produce
wide \ovi{} absorption with profiles that can be confused with the
Voigt profiles of DLAs. We exclude all the DLAs found in
the spectra of quasars with any positive Balnicity Index, as listed in
the DR12Q catalogue, leaving 23,342 DLAs.

  {\it Fourth cut: DLAs close to the \lya{} emission line}. All
systems within a velocity separation of $v_c < 5000\kms$ from the quasar
redshift are eliminated.
This condition is equivalent to requiring $\lambda_{r}\ge1195.39\angs$,
where $\lambda_{r}$ is the quasar rest-frame wavelength at which the
DLA absorption line is centered.
This reduces our sample to 21,408 DLAs.

 {\it Fifth cut: DLAs close to the OVI emission line}.
An excess of DLAs with $1005\angs < \lambda_{r} < 1037\angs$ was
found in FR12, likely caused by BAL
contamination. Removing all DLAs with $\lambda_{r}$ in this
interval reduces our sample to 19,655 DLAs. 

  {\it Sixth cut: DLAs in the Ly$\beta$ forest}. All the systems
blueward of the \lyb{} emission line are removed. This is done because,
as found in \cite{Mas-Ribas+2017}, a small fraction of the DLAs
detected bluewards of the \lyb{} line are in fact \lyb{} absorption
features for which the \lya{} line is not properly identified, and are
then confused with the \lya{} line of a DLA with the method of
\cite{Noterdaeme+2009}. This cut causes a considerable further reduction
of our sample to 13,734 DLAs. 

 The final sample contains a total of 13,734 DLAs. We emphasize
again that the purity of the sample is more important than its
completeness. However, we understand that the fourth, fifth, and
especially the sixth cut exclude an important amount of DLAs,
most of which will be true DLAs. To analyse the importance of
these cuts in the final measurement, different DLA samples are
studied in this work. We label the final sample considering all cuts as
dataset A, and the final sample considering only the cuts that are
most useful to remove contaminants (i.e., not applying cuts 4 to 6) as
C1. Finally, we label dataset C2 to be the sample resulting from the
application of all constraints save the sixth. In this sample, the same
cuts as in FR12 are applied, allowing for a more direct comparison. The
properties of the three datasets are summarized in
table \ref{ta DLA: sample properties}.

\begin{figure*}
	\centering
	\includegraphics[width=0.8\textwidth]{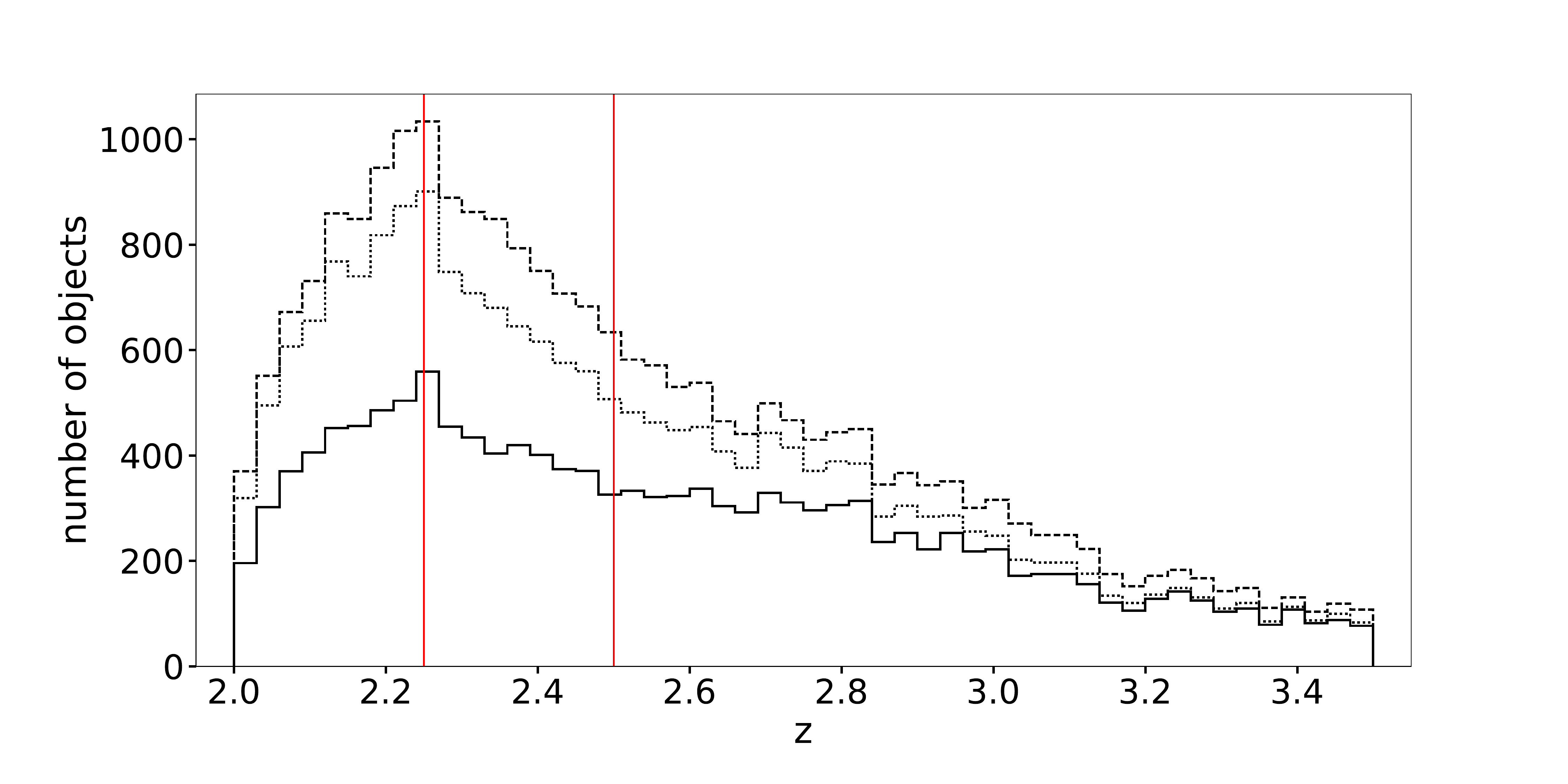}\\
	\includegraphics[width=0.8\textwidth]{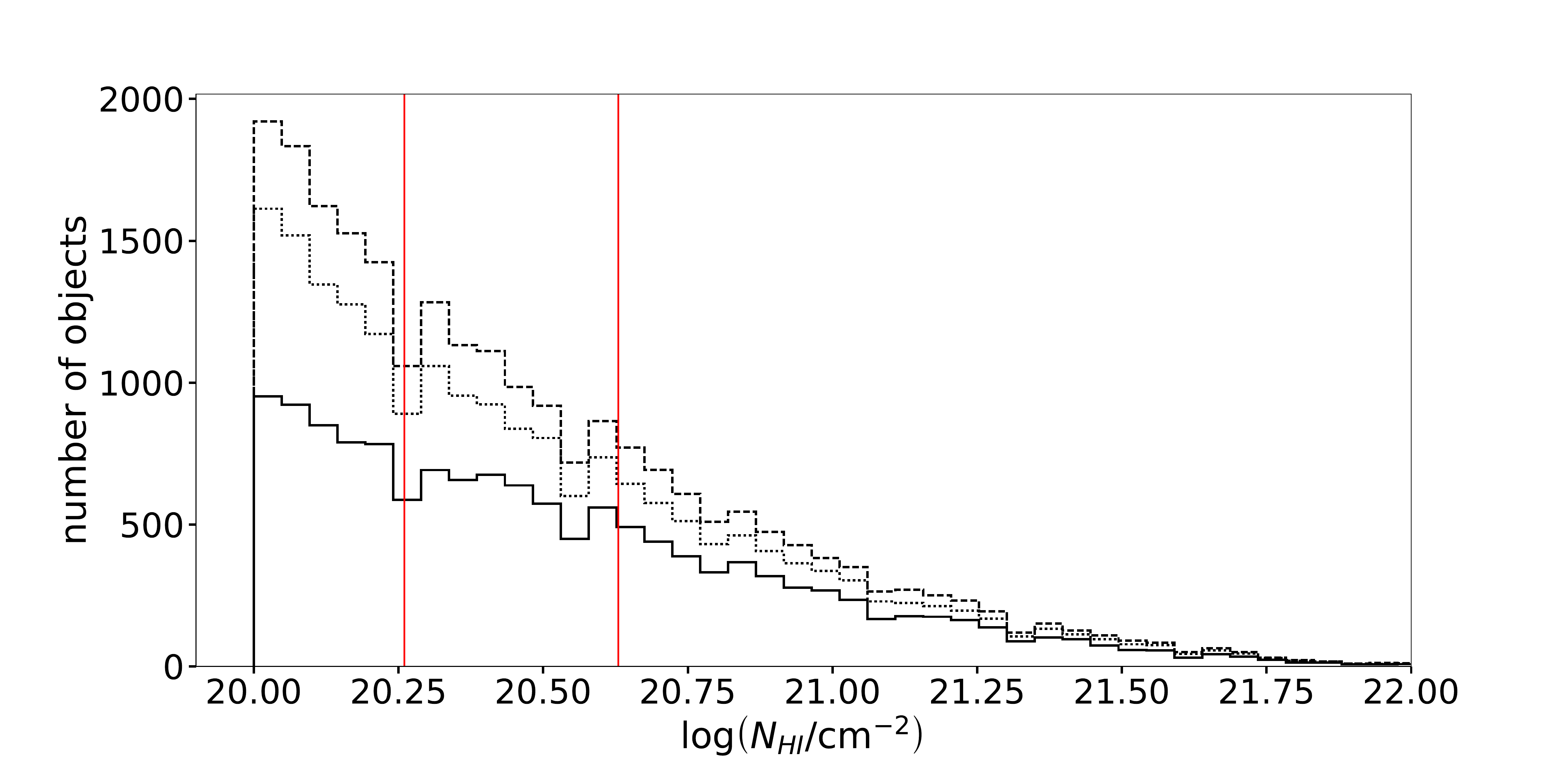}
	\caption{Distribution of the DLAs in samples A (solid),
C1 (dashed), and C2 (dotted) in redshift (top panel) and column density
(bottom panel). Solid red lines show the bins used to construct the
sub-samples (see table \ref{ta DLA: sample properties}) }
	\label{fig DLA: dla hist}
\end{figure*}

 We separate dataset A in bins of the DLA redshift and column
density. The bins are chosen in dataset A to obtain sub-sambles with
similar signal-to-noise ratio in the measured cross-correlation. We
label the redshift sub-samples Z1, Z2, and Z3, and the column density
sub-samples N1, N2, and N3, with properties listed in table
\ref{ta DLA: sample properties}. Figure \ref{fig DLA: dla hist} shows the
distribution of the total DLA sample in redshift and column
density. The bins used to define the sub-samples Z1 to Z3 and N1 to N3
are indicated as red solid lines, and are given in table
\ref{ta DLA: sample properties}.

\begin{table*}
	\centering
	\begin{tabular}{clc}
		\toprule
		Name & Description & Number of DLAs \\
		\midrule
		A	& full DLA sample & 13,734\\
		C1  & full DLA sample excluding cuts 4, 5 and 6 & 23,342\\
		C2	& full DLA sample excluding cut 6	&	19,655\smallskip\\
		Z1	& DLAs with $z_{\rm DLA}<2.25$ 	& 3,348\\
		Z2	& DLAs with $2.25\le z_{\rm DLA}<2.5$ 	& 3,455\\
		Z3	& DLAs with $2.5\le z_{\rm DLA}$		& 6,931\smallskip\\ 
		N1	& DLAs with $\log\left(N_{HI}/{\rm cm^{-2}}\right)<20.26$ & 4,448\\
		N2	& DLAs with $20.26\le\log\left(N_{HI}/{\rm cm^{-2}}\right)<20.63$ & 4,683\\
		N3	& DLAs with $20.63\le\log\left(N_{HI}/{\rm cm^{-2}}\right)$ & 4,603\\
		\bottomrule
	\end{tabular}
	\caption{Summary of DLA samples A, C1 and C2, and
sub-samples of sample A with the indicated redshift and column density bins.}
	\label{ta DLA: sample properties}
\end{table*}

\subsection{\texorpdfstring{\lya{}}{Lya} Sample} \label{subs DLA: lya Sample}

 For the \lya{} sample we use the same set of \lya\ spectra of DR12
as in \cite{Busca+2013}, with a total of 157,922 spectra containing over
27 million \lya{} pixels. We use their {\it analysis pixels} which are
the average of every three pixels of the actual co-added spectra, because
our cross-correlation measurements do not depend on small-scale
variations and this saves computational time. Throughout the rest of
this paper, {\it pixel} refers to analysis pixels unless otherwise
stated. The effective width of these pixels is
$(\Delta \lambda/\lambda) c = 207\kms$. 

 The \lya{} transmission fluctuation at every pixel $i$ with wavelength
$\lambda_i$ and measured flux $f_i$ is defined as
\begin{equation}
	\label{eq: delta definiton}
	\delta_{i} = \frac{f_{i}}
 {C_{q}\left(\lambda_{i}\right)\overline{F}\left(z_{i}\right)} - 1 ~.
\end{equation}
Here, $C_{q}\left(\lambda\right)$ is the quasar continuum (or
unabsorbed flux), and $\overline{F}\left(z\right)$ is the mean
transmitted fraction at the \lya\ absorber redshift. The pixel redshift
is $z_{i} = \lambda_{i}/\lambda_{\rm Ly\alpha} - 1$. 
We use the quasar continuum designated as {\it method 1} in
\cite{Busca+2013}, which assumes a universal shape of the quasar
rest-frame continua except for a multiplicative factor that is linear
in wavelength which allows for a variable slope of the quasar
continuum that is fitted to the \lya{} forest region. We refer the reader
to \cite{Busca+2013} for a more detailed description of this method.

  An important difference relative to FR12 is that we correct the \lya\
forest transmission for the DLAs that are identified in the DR12 DLA
catalogue that we use. If this is not done, the DLA-\lya\
cross-correlation includes a component that is caused by the DLA
autocorrelation, owing to the contribution from DLAs to the \lya\
absorption spectra. We apply this correction in the same way as
\cite{Bautista+2017}: for every DLA in the catalogue, we compute its
absorption Voigt profile and we eliminate any pixels in which the
computed DLA transmission is less than $0.8$. We then correct all other
pixels in the spectrum by dividing the measured transmission by the
computed DLA transmission. This eliminates only the detected DLAs, so
the absorption of any undetected systems remains (these systems are
generally of low column density in low signal-to-noise spectra). We
discuss how this is modelled in section \ref{sec DLA: Model}.

%
\section{Cross-correlation}  \label{sec DLA: Cross-correlation}

\subsection{Estimator for the cross-correlation}

 In this section we describe the method used to compute the
cross-correlation of DLAs and the \lya\ transmission fluctuation
$\delta_i$, and its covariance matrix. The method is similar to that
used in FR12, although the broadband uncertainties arising from the
continuum fitting of quasar spectra are treated in a different way. FR12
used a simple estimator of the cross-correlation $\xi$ as a function
of the parallel and perpendicular components of the separation vector
{\bf r} between a DLA and a \lya{} pixel, given by
\begin{equation}
	\label{eq DLA: xi}
	\tilde{\xi}^{A} = \frac{\sum_{i\in A}w_{i}\delta_{i}}{\sum_{i\in A}w_{i}} ~,
\end{equation}
where the sum is over all DLAs and over all the pixels $i$ located
within a bin A of the separation {\bf r} from a DLA, and the weights $w_i$
are defined to optimize the accuracy of the measurement of $\xi^{A}$. They
then performed a {\it mean transmission correction} to compensate for
the effects of the quasar
continuum fitting. Note that a given pixel $i$ of the \lya{} forest appears
as many times in the sum in equation \ref{eq DLA: xi} as
there are DLAs at a separation from the pixel within bin A.

  We adopt a different approach, following the one used in \cite{Bautista+2017}. We present a brief
description of the method here, and a more extended and detailed
explanation in appendix \ref{sec DLA: proj}. The goal is to remove from
the cross-correlation the part that is strongly affected by systematics
related to the continuum fit, by using an adequate projector. The
effect of this projector can then be taken into account in the
modelling, eliminating the need for the mean transmission correction.
Our assumption is that the measured \lya{} transmission fluctuation
$\delta^{(m)}$ differs from the true one, $\delta^{(t)}$, by a linear
additive function,
\begin{equation}
	\label{eq DLA: hypothesis 2}
	\delta_{i}^{(m)} = \delta_{i}^{(t)}+a+b \log \lambda_{i} ~,
\end{equation}
where $a$ and $b$ are unknown for each forest. That is, we assume that a
linear approximation to the continuum in the region of the \lya{} forest
adequately describes the effect of all the systematic calibration errors
in the observed spectrum and of having fitted a continuum to it.
Although in this paper we use this linear expansion in $\log\lambda_i$,
this method works the same way if the linear fit is assumed in
$\lambda_i$ instead.
We define a projector $P_f$ for each forest $f$ that removes this
unknown part by subtracting a weighted linear regression to the forest,
so that the projected measured and true fluctuations are equal:
\begin{equation}
	\label{eq DLA: condition}
         \delta_i \equiv \sum_{j\in f}P_{f,ij} \delta_j^{(m)} =
 \sum_{j\in f} P_{f,ij} \delta_j^{(t)} ~.
\end{equation}
The sums are over all pixels $j$ that belong to the same forest $f$ as
pixel $i$. From this point on, we use $\delta_i$ to mean the projected
transmission fluctuation, after subtracting the weighted linear
regression by applying the projector $P_f$. A more detailed derivation
of the equation for this projector is given in appendix \ref{sec DLA: proj}.
The cross-correlation in bin $A$ is then expressed by exactly the same equation \ref{eq DLA: xi}, except that now $\delta_i$ is
understood to have been projected.

 In general, this projector can introduce an artificial non-vanishing
correlation at large scales, arising from a mean value of
$\delta = P_f \delta^{(m)}$ at a given redshift that is not equal to
zero, because only the mean value of $\delta^{(m)}$ in narrow redshift
bins was initially required to be zero. We solve this by
computing the cross-correlation of the mean transmission value,
$\bar\delta_i$, at the redshift $z_i$ of pixel $i$, designated as
$\tilde{\xi}^{A}_{\rm sky}$, using the same equation
\ref{eq DLA: xi}, and then subtracting it as a correction. The final
cross-correlation is
\begin{equation}
	\label{eq DLA: xi corrected}
	\xi^{A} = \tilde{\xi}^{A} - \tilde{\xi}^{A}_{\rm sky}~.
\end{equation}
For the cross-correlation between DLAs and the \lya{} forest,
this correction is negligible at our current level of precision, but this needs not be the
case in general.

\subsection{Covariance matrix} \label{subs DLA: Covariance matrix}

 The covariance of the cross-correlation at two bins $A$ and $B$ is equal to
\begin{multline}
	\label{eq DLA: covariance matrix}
	C^{AB} \equiv \left<\xi^{A}\xi^{B}\right>-\left<\xi^{A}\right> \left<\xi^{B}\right> =
	\\=\frac{1}{S^{AB}} \sum_{i\in A} \sum_{j\in B} w_iw_j\, \zeta_{ij} ~,
\end{multline}
where $\zeta_{ij}$ is the \lya{} forest autocorrelation of the values of
$\delta$ at pixels $i$ and $j$, and each of the two sums are again
understood to be over all \lya{} forest pixels and all the DLAs
at separations within the bins $A$ or $B$. The normalization factor is
\begin{equation}
	\label{eq DLA: covariance matrix norm factor}
	S^{AB} = \sum_{i\in A} \sum_{j \in B} w_i w_{j} ~.
\end{equation}

  As discussed in \cite{Font-Ribera+2012}, there are three main
contributions to the correlation $\zeta_{ij}$. First, there is a noise
component that we assume to be uncorrelated among different pixels, and
is therefore present only for $i=j$. This contribution arises from the
fact the same \lya{} pixel contributes several times to the evaluation of
$\xi$ at different bins when it is paired with different
DLAs. Second, there is a contribution produced by continuum
fitting errors inducing a correlation among pixels in the same forest.
Finally, different \lya{} pixels are intrinsically correlated due to
the physical \lya{} forest autocorrelation. This entire autocorrelation
$\zeta_{ij}$ can be measured directly from the data, but in practice it
is computationally expensive to compute the covariance matrix taking
into account the correlation among pixels in different forests out to
a large transverse separation, because of the large number of
DLA-\lya{} pixels pairs-of-pairs involved in the sum of equation
\ref{eq DLA: covariance matrix}. In this work we neglect the
contribution to the covariance matrix of pixels in different forests. 
We find, however, that it is important to measure the change of $\zeta$
with redshift.
Once we restrict this autocorrelation to pixel pairs on a single
forest, $\zeta_{ij}$ can be expressed as a function of the redshift $z$
and the separation $n=j-i$ in number of pixels between $j$ and $i$
along the line of sight,
\begin{equation}
	\label{eq DLA: covariance matrix 3}
	\zeta(z, n)=\frac{
 \sum_{i, z_i = z} w_i w_{i+n}\delta_{i} \delta_{i+n}}
 {\sum_{i, z_i = z} w_{i}w_{i+n}} ~,
\end{equation}
where the sum is over all pixels $i$ which have redshift $z_{i}=z$, and the $\delta_i$ are as usual the projected transmission
fluctuations.

  We compute this autocorrelation in redshift bins of width
$\Delta z=0.0037$ for $n$ up to 5. We have checked that further
increasing the maximum value of $n$ does not modify the recovered
covariance matrix, while it increases the computational time.

\subsection{Distortion matrix} \label{subs DLA: Distortion matrix}

 Having applied the projection to the data to eliminate the most important
continuum fit systematics, we need to correct the model we fit to include
the effect of this projection. The mixing of the $\delta$ variables in
the same forest due to this projection implies that the projected
cross-correlation in bin A, $\xi_p^{A}$, is related to the
model cross-correlation $\xi_m$ by a distortion matrix $D$,
\begin{equation}
	\xi_p^{A} = \sum_{B}D^{AB}\xi_m^{B} ~.
\end{equation}
The distortion matrix element $D^{AB}$ relates the projected
cross-correlation in bin $A$ to the model cross-correlation at all
bins $B$, and can be directly computed from the quasar positions in
the survey and the redshift range of each forest being used, with the
same method that was used in \cite{Bautista+2017}. The resulting
$\xi_p$ is the one that is compared to the projected measured
cross-correlation $\xi$ in equation \ref{eq DLA: xi corrected} to fit
any given model. The detailed way we compute the distortion matrix is
explained in appendix \ref{subs DLA: distortion matrix appendix}.

\subsection{Bootstrap errors}\label{subs DLA: bootstrap}

 The errors obtained when computing the covariance matrix rely on the
validity of the approximations we have made. One of the most important
approximations is that we include DLA-\lya{} forest pairs-of-pairs
only when the two \lya\ pixels are in the same forest. We also neglect
errors associated with spectral calibration, which are
difficult to model reliably for including them in a direct calculation
of the covariance matrix. It is therefore important to test the validity
of our errors by computing them alternatively using the bootstrap method.

  We divide the survey into sub-samples using the plate number of the
observations. Each DLA-\lya{} pair is always assigned to the plate that
the \lya{} pixel belongs to. Using the 2,400 regions defined by the
plates, a total of 100 bootstrap samples are generated. We compute the
cross-correlation for each of these bootstrap samples and then we fit
our model (see section \ref{sec DLA: Model}), modified by the distortion matrix
mentioned above and using the covariance matrix to compute the $\chi^2$.
The distortion and covariance matrices are computed for the whole sample
and not modified for each of the 100 bootstrap resamplings. The
bootstrap errors of model parameters are computed in the standard way,
equal to the dispersion of the best-fit parameter values obtained in
the bootstrap samples.

%
\section{Fitting the cross-correlation}  \label{sec DLA: Model}

  This section describes the linear theory model that is used to fit the
measured DLA-\lya{} forest cross-correlation. All the actual fits
are computed with the publicly available fitting code {\it baofit}
\citep[][see \protect\url{http://darkmatter.ps.uci.edu/wiki/DeepZot/Baofit}]{Kirkby+2013}.

  In the limit of large scales linear theory predicts the form of the
cross-correlation of any two tracers of the large-scale mass-density
fluctuations. The limit of large scales is broadly expected to apply
when the relative mass-density fluctuation is small compared to unity
at the redshift of our observations, but the precision at which linear
theory is reliable depends on the tracer. For the \lya{} forest, linear
theory often works surprisingly well because the transformation from
optical depth to transmission fraction suppresses the contribution from
highly overdense regions, which develop the largest non-linearities, to
the measured correlations.

  In real space, any biased tracer should have the same linear
fluctuations as the mass-density, except for a linear biased factor.
For example, the \lya{} forest transmission fluctuation at any pixel $i$,
after being smoothed three-dimensionally over a large scale, would
simply be related to the mass fluctuation $\delta_m$ smoothed in the
same way by $\delta_i = b_{\rm Ly\alpha} \delta_m$, if the effects of peculiar
velocities were somehow eliminated. In Fourier space, the same relation
holds for the Fourier modes. However, observations can only be done in
redshift space, where peculiar velocity gradients enhance the amplitude
of each Fourier mode according to the expression found by
\cite{Kaiser1987},
\begin{equation}
 \delta_i = b_{\rm Ly\alpha} \left(1+\beta_{\rm Ly\alpha} \mu_k^2\right) \, \delta_m ~,
\end{equation}
where $\beta_{\rm Ly\alpha}$ is the redshift distortion parameter, and
$\mu_{k}$ is the cosine of the angle between the Fourier mode vector and
the line of sight. The density fluctuations of DLAs also have
their own bias and redshift distortion parameter, and the linear
cross-power spectrum of the two types of objects is equal to
\begin{multline}
	\label{eq DLA: PS model}
	P_{\rm DLA,Ly\alpha}\left({\bf k}, z\right) = b_{\rm DLA}
 \left(1+\beta_{\rm DLA}\mu_k^2 \right) \times \\ \times b_{\rm Ly\alpha}
 \left(1+\beta_{\rm Ly\alpha}\mu_k^2 \right) \,
 P_L(k, z)\, G({\bf k})\, S(k_{\parallel}) ~,
\end{multline}
where $P_L(k, z)$ is the linear matter power spectrum. We have
introduced also two smoothing functions of the cross-correlation, which
are multiplicative functions in Fourier space: $S(k_{\parallel})$
accounts for the spectrograph resolution and binning of the \lya{}
forest spectra, and $G({\bf k})$ accounts for the binning used to
compute the cross-correlation function. For the calculation of the
linear power spectrum $P_L$, BAOFIT uses templates that were computed
for our specific cosmology using CAMB at the reference redshift
$z_{\rm ref}=2.3$ \citep{Kirkby+2013}.

  The DLA-\lya{} cross-power in equation \ref{eq DLA: PS model}
depends only on the product of the two bias factors $b_{\rm DLA}$ and
$b_{\rm Ly\alpha}$. We can therefore infer the value of one of the
bias factors only if the other one, as well as the normalization of
$P_L$, is independently constrained. The two redshift distortion bias
factors have effects that are also difficult to separate, and only one
of them can be measured in practice from the shape of the
cross-correlation in redshift space. Previous analyses of the
BOSS Collaboration
\citep[see e.g.][]{Blomqvist+2015, Delubac+2015, Bautista+2017} have
studied in detail the \lya{} forest autocorrelation and obtained
constraints on the \lya{} forest bias factors. We use the values listed
in table 3 of \cite{Bautista+2017}: $\beta_{\alpha}=1.663\pm0.085$ and
$b_{\alpha}\left(1+\beta_{\alpha}\right)=-0.325\pm0.004$, at a reference
redshift $z_{\rm ref}=2.3$. We fix these two \lya{} forest parameters to
their mean values from this measurement. The errors and modeling
uncertainties of the \lya{} forest bias factors obtained in this way
introduce systematic errors in our derived DLA bias factor, which are
discussed in detail in section \ref{subs DLA: lya}.

  We do not include in our model any additive broadband function 
to measure the form of the cross-correlation, which can arise from
spectral calibration systematics and continuum fitting in the \lya{}
forest region, and have been used in previous studies of the BOSS
\lya{} data where the focus was in measuring the narrow-band feature
of the Baryon Acoustic Oscillation peak in the correlation function
\citep[e.g.][]{Font-Ribera+2014, Blomqvist+2015, Delubac+2015, Bautista+2017}

  The model is evaluated at the mean values of the parallel and
perpendicular components of the separation vector, $r_{\parallel}$ and
$r_{\perp}$, for each of the bins of the measured cross-correlation, and
at the mean redshift of our sample. For the evolution with redshift, we
assume that $b_\alpha \propto (1+z)^{2.9}$, and that $b_{\rm DLA}$ and
the redshift distortion parameters $\beta_\alpha$ and $\beta_{\rm DLA}$
are constant. This evolution of $b_\alpha$ follows that
measured from previous \lya{} autocorrelation studies
\citep[e.g.,][]{McDonald+2006}, and we shall see below that a constant
$b_{\rm DLA}$ with redshift is consistent with our results. Including
the linear growth factor, this implies that the amplitude of the
cross-power spectrum in equation (\ref{eq DLA: PS model}) evolves
approximately as $(1+z)^{0.9}$. We fix
$\beta_{\rm DLA}b_{\rm DLA}=f\left(\Omega\right)=0.968897$, assuming
that there is no peculiar velocity gradient bias for DLAs.

  The term $G\left({\bf k}\right) = G_{\parallel}\left(k_{\parallel}\right)\,
G_{\perp}\left(k_{\perp}\right)$ corrects for the binning in
$r_{\parallel}$ and $r_{\perp}$ which averages the cross-correlation
over a bin. We use
$G_{\parallel}\left(k_{\parallel}\right) =
 {\rm sinc}^{2}\left(\Delta_{\parallel}k_{\parallel}/2\right)$ and
$G_{\perp}\left(k_{\perp}\right) =
 {\rm sinc}^{2}\left(\Delta_{\perp}k_{\perp}/2\right)$, as in
\cite{Bautista+2017}, where $\Delta_{\parallel}$ and $\Delta_{\perp}$
are the bin sizes. In this work they are both equal to $2\hmpc$. 
We also correct for the spectrometer resolution and for the averaging of
the three spectrometer pixels into analysis pixels, by approximating the
convolution of a Gaussian and a top-hat as a new Gaussian,
$S\left(k_{\parallel}\right) = \exp[ - k_\parallel^2/(2\sigma_S^2)]$.
The contribution to the variance $\sigma_S$ from the Point Spread Function (PSF) of the
BOSS instrument is set to $\sigma_{PSF}=0.61\hmpc$ (in comoving units),
which corresponds to a full-width half-maximum
$R=(\Delta\lambda/\lambda)^{-1}=2000$ at the reference redshift
$z_{\rm ref}=2.3$. The averaging of three spectrometer pixels, which
have a top-hat full width
$\Delta\lambda/\lambda=3\, \ln(10) \times 10^{-4}= 6.91\times 10^{-4}$
\citep{Busca+2013}, contributes an additional dispersion
$\sigma_{p}=cH^{-1}(z)(1+z)\, \Delta\lambda/\lambda/\sqrt{12} = 0.57\hmpc$,
also in comoving units. The overall dispersion is
$\sigma_{S} =\sqrt{\sigma_{PSF}^{2}+\sigma_{p}^{2}} = 0.83\hmpc$.

 In section \ref{sec DLA: Sample data}, we explained how the absorption
profiles of DLAs also contribute to the \lya{} forest transmission and
therefore to the measured DLA-\lya{} cross-correlation.
While the detected DLAs are corrected, many DLAs remain undetected in
low signal-to-noise spectra,
and all the absorption systems with column densities $N_{\rm HI} <
 10^{20}\cm^{-2}$, which are not considered to be DLAs but also have
damped absorption wings, contribute to the \lya{} transmission. These
systems cannot be removed or corrected directly in the data, and
therefore their effect needs to be corrected from the measured
cross-correlation.

 In general, the measured cross-correlation is the sum of the
cross-correlations of DLAs with several populations of objects
that contribute to the absorption in the \lya{} forest spectra. The
population of hydrogen absorbers including unidentified DLAs and systems
of lower column density that have significant damped wings is designated
as high-column density systems, or HCDs. In addition to these, some
metal lines with wavelengths close to the \lya{} line can also contribute
significantly to the cross-correlation, and were modelled in
\cite{Bautista+2017}. We ignore these metal lines here, because the
signal-to-noise ratio of the DLA-\lya{} cross-correlation is smaller than
the \lya{} auto-correlation, and the effect of metal lines is not clearly
discernible in our results; this is further addressed in section
\ref{subs DLA:model dependence}. We include only the HCDs as an additive
contamination,
\begin{equation}
	\label{eq: metals+HCD}
	\xi^{A}_{\rm obs} = \xi^{A}_{{\rm DLA-Ly}\alpha } +
	  \xi^{A}_{\rm DLA-HCD} ~,
\end{equation}
where $\xi^A_{\rm DLA-Ly\alpha }$ is the Fourier Transform of the power
spectrum in equation \ref{eq DLA: PS model}).
The cross-correlation with HCDs is assumed to be the Fourier transform
of the same linear theory form of the cross-power as our model for DLAs:
\begin{multline}
	\label{eq: xiHCD}
  P_{\rm DLA-HCD} = b_{\rm DLA}\, b_{\rm HCD}\,
 (1+\beta_{\rm DLA} \mu_k^2)\times \\ \times (1+\beta_{\rm HCD} \mu_k^2)
P_L(k,z)
 F_{\rm HCD}(k_\parallel ) ~,
\end{multline}
where the function $F_{\rm HCD}(k_\parallel )$ is introduced to
approximately model the average wavelength profile of HCDs, and is set
to
\begin{equation}
F_{\rm HCD}\left(k_{\parallel}\right) =
 \sin (L_{\rm HCD} k_{\parallel} )/ (L_{\rm HCD} k_{\parallel} ) ~,
\end{equation}
where $L_{\rm HCD}$ is a parameter that reflects the width of the
absorption wings of HCDs. We use the values found in
\cite{Bautista+2017} to fit the observed \lya{} autocorrelation,
listed in their Table 3: $b_{\rm HCD}= -0.0288$,
$\beta_{\rm HCD}=0.681$, and $L_{\rm HCD}=24.34\hmpc $. The bias
$b_{\rm HCD}$ is assumed to evolve with redshift in the same way as
the \lya{} forest bias, $b_{\rm HCD}\propto (1+z)^{2.9}$, for reasons of
computational efficiency (this evolution makes very little difference
to the computed effect of HCDs; the value given above is at the
reference redshift $z_{\rm ref}=2.3$), and the other two parameters are
assumed to be independent of redshift.

%
\section{Results}  \label{sec DLA: Results}

 We have measured the cross-correlation for all the samples listed in
table \ref{ta DLA: sample properties}, with bin sizes
$\Delta_{\parallel}=\Delta_{\perp}=2\hmpc$, out to a maximum separation
of $80\hmpc$ both in the parallel and perpencicular directions. In
this section, all the model parameters as described in the previous
section are fixed, and we fit only $b_{\rm DLA}$. Note that the DLA
redshift distortion parameter, $\beta_{\rm DLA}= f(\Omega)/b_{\rm DLA}$,
also varies with $b_{\rm DLA}$; this is, however, a small effect,
because the redshift distortions of the cross-correlation are dominated
by $\beta_{\rm Ly\alpha }$, and variations of $\beta_{\rm DLA}$ in all the
results we present are small. Neglecting the variation of
$\beta_{\rm DLA}$, fitting the bias $b_{\rm DLA}$ is equivalent to
fitting the amplitude of our cross-correlation model with a fixed shape
to the data, and this amplitude is proportional to $b_{\rm DLA} b_{\rm Ly\alpha}\sigma_8^2$, where $\sigma_8^2$ is the standard quantity to express
the normalization of the power spectrum $P_L$.

  Our cross-correlation model assumes linear theory, and therefore we
exclude bins at a small value of
$r= (r_{\parallel}^{2}+r_{\perp}^{2})^{1/2}$ in the fits to reduce the
impact of non-linearities on our result. For each sample we perform two
fits, a conservative one that excludes bins with $r< r_{\rm min}= 10\hmpc$,
and a more generous one excluding only bins with $r< r_{\rm min}= 5\hmpc$.
In addition, all fits exclude bins with $r > 90\hmpc$ to better define
the radius range of our measurements; we shall see that the
cross-correlation signal is not clearly detected beyond $r \gtrsim 60 \hmpc$.

\subsection{Measured cross-correlation and DLA bias} \label{subs DLA: bias}

  The measured values of $b_{\rm DLA}$ are summarized in
table \ref{ta DLA: biases}. Results for the A, C1 and C2 samples (see
section \ref{sec DLA: Sample data}) are presented in section
\ref{subs DLA: bias}, the redshift and column density dependence are
explored in section \ref{subs DLA: bias z} using the subsamples Z1 to Z3
and N1 to N3, and the scale dependence of the bias factor is
investigated in section \ref{subs DLA: bias r}.

\begin{table*}
	\centering
	\begin{tabular}{ccccc}
		\toprule
		& \multicolumn{2}{c}{$r_{\rm min}=10\hmpc$} & \multicolumn{2}{c}{$r_{\rm min}=5\hmpc$} \\
		Dataset & $b_{\rm DLA}$ & $\chi^{2} (d.o.f.)$ & $b_{\rm DLA}$ & $\chi^{2} (d.o.f.)$ \\
		\midrule
		A	& $2.00\pm0.15(0.19)$ & 2,817.43  (2,864-1) & $2.06\pm0.11(0.14)$ & 2,854.08 (2,896-1)\\
		C1  & $1.93\pm0.11(0.13)$ & 3,019.44 (2,864-1) & $1.97\pm0.08(0.10)$ & 3,065.79 (2,896-1)\\
	    C2	& $1.97\pm0.12(0.14)$ & 2,911.86 (2,864-1) & $1.99\pm0.09(0.11)$ & 2,950.26 (2,896-1)\smallskip\\
		Z1  & $2.40\pm0.24(0.31)$ & 2,906.85 (2,864-1) & $2.36\pm0.17(0.21)$ & 2,936.71 (2,896-1) \\
		Z2  & $1.39\pm0.25(0.29)$ & 2,875.71 (2,864-1) & $1.90\pm0.18(0.21)$ & 2,944.79 (2,896-1)\\
		Z3  & $2.27\pm0.29(0.31)$ & 2,807.96 (2,864-1) & $1.92\pm0.20(0.23)$ & 2,855.82 (2,896-1)\smallskip\\
		N1  & $2.05\pm0.26(0.32)$ & 2,844.55 (2,864-1) & $2.09\pm0.19(0.26)$ & 2,869.06 (2,896-1) \\
		N2  & $2.33\pm0.26(0.32)$ & 2,929.53 (2,864-1) & $2.17\pm0.18(0.23)$ & 2,955.24 (2,896-1) \\
		N3  & $1.60\pm0.26(0.28)$ & 2,847.15 (2,864-1) & $1.92\pm0.18(0.20)$ & 2,891.66 (2,896-1)\\
		\bottomrule
	\end{tabular}
	\caption{Summary of the fitted $b_{\rm DLA}$ for each
DLA subsample, with the values of $\chi^2$ for the fits with
only one free parameter. The values of the bias are given at the reference
redshift $z_{\rm ref}=2.3$. Errors are obtained from our computed covariance
matrix, and also using the bootstrap method (shown in parenthesis).
See table \ref{ta DLA: sample properties} for the subsample definitions.}
	\label{ta DLA: biases}
\end{table*}
 
 Our fiducial result to which we refer for all comparisons is the fit
to sample A, which yields $b_{\rm DLA} = 2.00\pm0.19$ for
$r_{\rm min}=10\hmpc$, and $b_{\rm DLA} = 2.06\pm0.14$ for
$r_{\rm min}=5\hmpc$, at a reference redshift $z_{\rm ref}=2.3$.
In general, the fits reported in table \ref{ta DLA: biases} have
covariance matrix errors lower than the errors derived by the bootstrap
technique. This is likely because the \lya\ transmission correlations in
different forests are neglected when computing the covariance matrix.
The bootstrap errors should therefore be considered as more reliable.

  The values of $\chi^2$ of the fit to the measured cross-correlation
indicate that our model is fully consistent with the data for sample A,
and marginally inconsistent at the $\sim 3-\sigma$ level for sample C1,
for both values of $r_{\rm min}$. This may be due to a contamination of
the signal introduced by false DLAs that appear near the \lya\ and OVI
emission lines of quasars, which are not removed in sample C1.

\begin{figure*}
	\centering
	\includegraphics[width=0.78\textwidth]{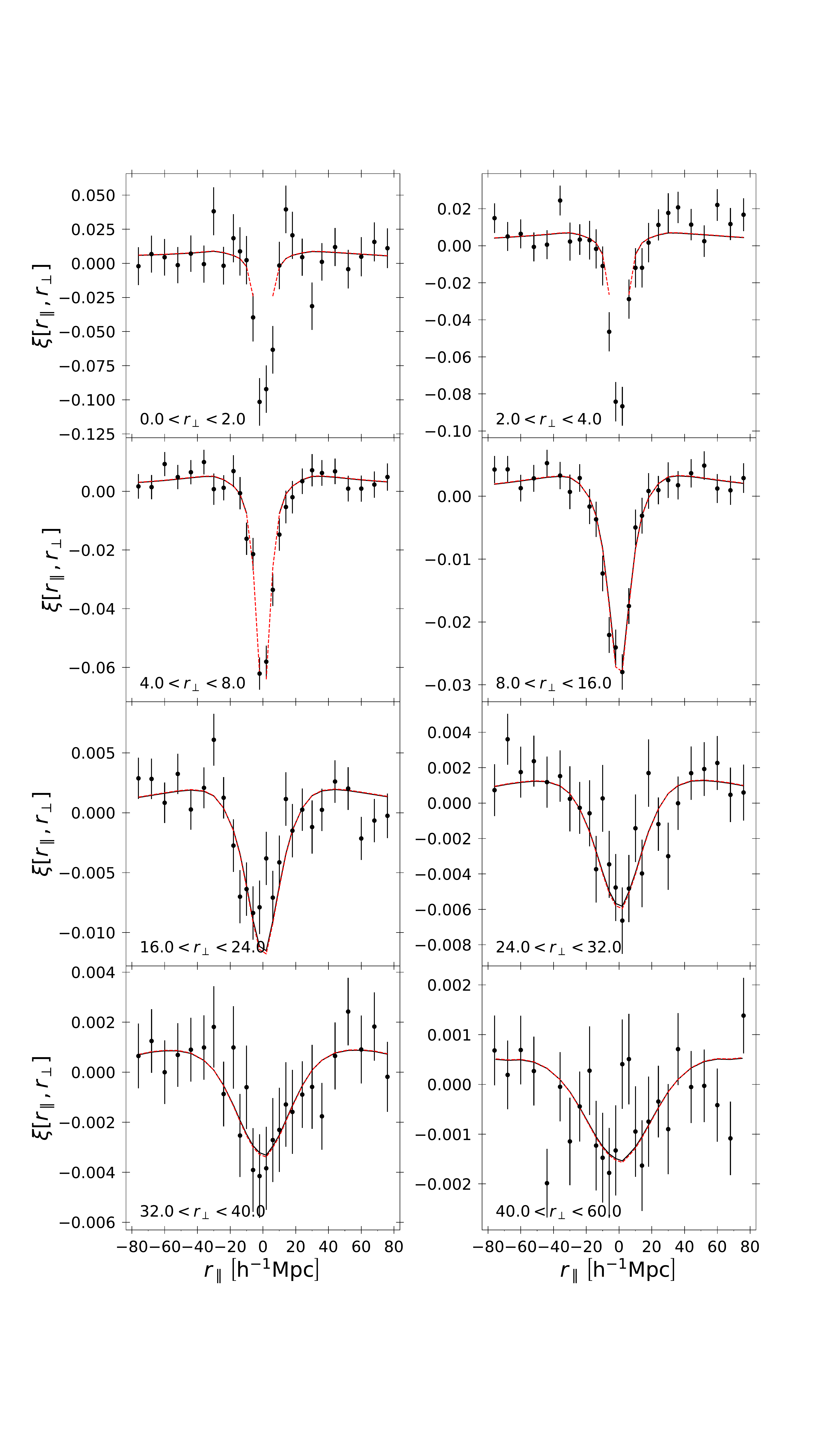}
	\vspace{-2.cm}
	\caption{Cross-correlation of DLAs and \lya{} forest as a
function of $r_{\parallel}$ for various bins in $r_{\perp}$, in
comoving $\hmpc$, for sample A. Black circles show the measured
cross-correlation for sample A. Solid black lines and dashed red lines
correspond to the best-fit model considering $r_{\rm min}=10\hmpc$ and
$r_{\rm min}=5\hmpc$, respectively, and are nearly equal and hard to
distinguish in the figure. Data and models have been rebinned to wider
bins than used in the analysis to plot this figure.}
	\label{fig DLA: cross_correlation}
\end{figure*}
 
\begin{figure*}
	\centering
	\includegraphics[width=\textwidth]{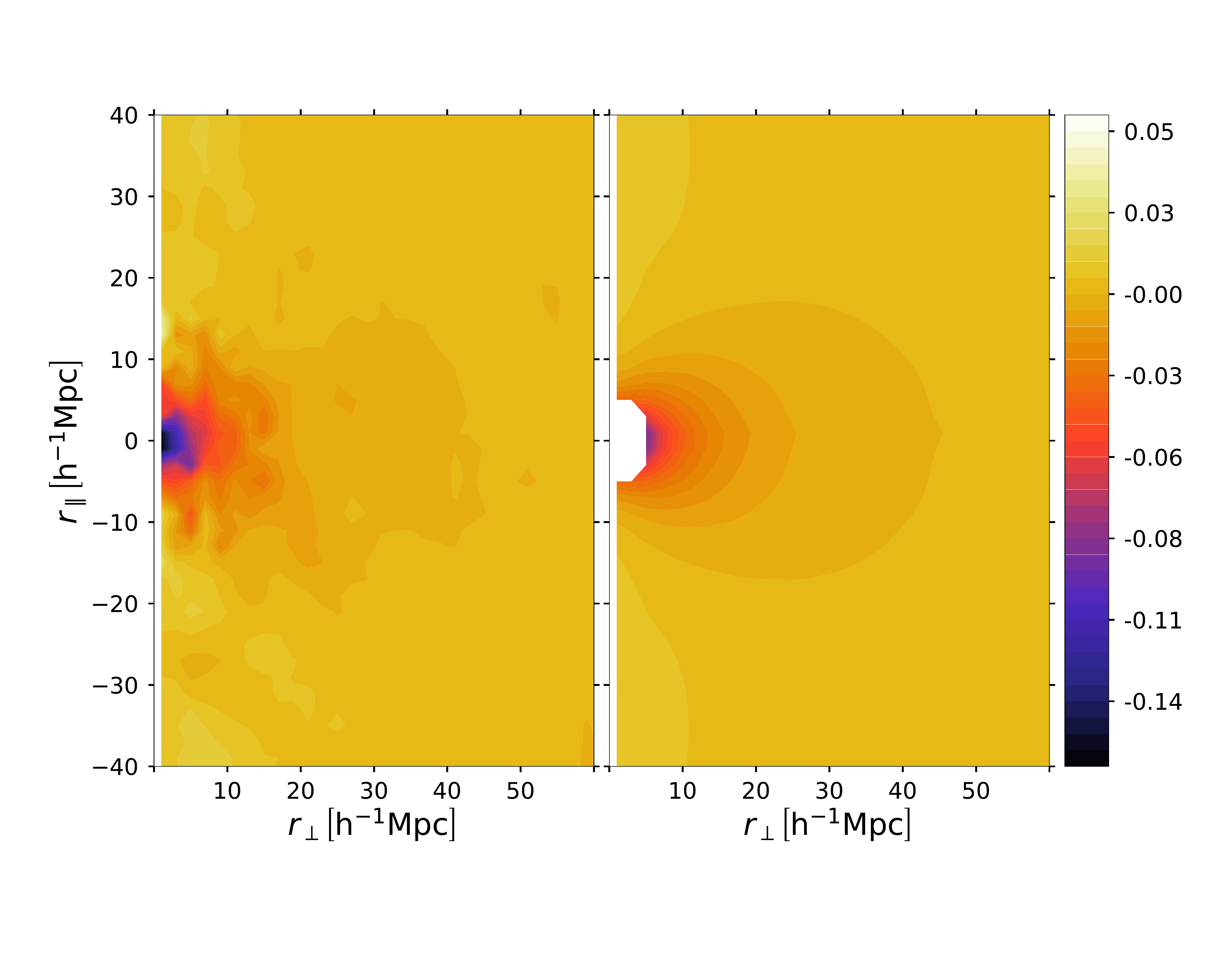}
	\caption{Smoothed contour plots of the measured
DLA-\lya{} cross-correlation (left, sample A) and best-fit
theoretical model considering bins with
$5\hmpc < r=(r_{\parallel}^2+r_\perp^{2})^{1/2} < 90\hmpc$ (right).}
	\label{fig DLA: cross_correlation contour}
\end{figure*}

 The results of the DLA-\lya{} cross-correlation as a function of
$r_{\parallel}$ are shown for various bins of $r_{\perp}$ in figure
\ref{fig DLA: cross_correlation}, for sample A. We have rebinned the
cross-correlation measurements into wider bins than the ones used for
computing the fits in both $r_\parallel $ and $r_\perp $, for display
purposes only, recomputing the plotted errors in
these wider bins using our covariance matrix. Results are shown only out
to $r_{\perp} = 60\hmpc$, even though our measured cross-correlation is
used in all the bins out to $r_{\perp}=80\hmpc$. These results are also
shown as a contour plot with smoothed contours in figure
\ref{fig DLA: cross_correlation contour} (left panel).

 The black solid and red dashed lines in figure
\ref{fig DLA: cross_correlation} are our best-fit models for
$r_{\rm min}=10\hmpc$ and $r_{\rm min}=5\hmpc$, respectively. In
practice the two curves are nearly identical (the difference in
$b_{\rm DLA}$ is only 3\%) and can hardly be distinguished. The
curves are not shown in bins at small $r$ that were not used for the
fit, although when the model is averaged into the wider bins for
plotting purposes, we include all bins even if they are not used in
the fit to facilitate a correct comparison with the data points.
The model for the case $r_{\rm min}=5\hmpc$ is also presented in a
contour format in figure \ref{fig DLA: cross_correlation contour}
(right panel). 

 When relaxing the cuts imposed in the sample A, we find that
$b_{\rm DLA}$ is slightly lower in samples C1 and C2. This may be partly
due to a decreased purity when we eliminate some of the cuts imposed on
sample A, although the differences are consistent with statistical
errors.

\subsection{Bias dependence on redshift and column density} \label{subs DLA: bias z}

\begin{figure*}
	\centering
	\includegraphics[width=\textwidth]{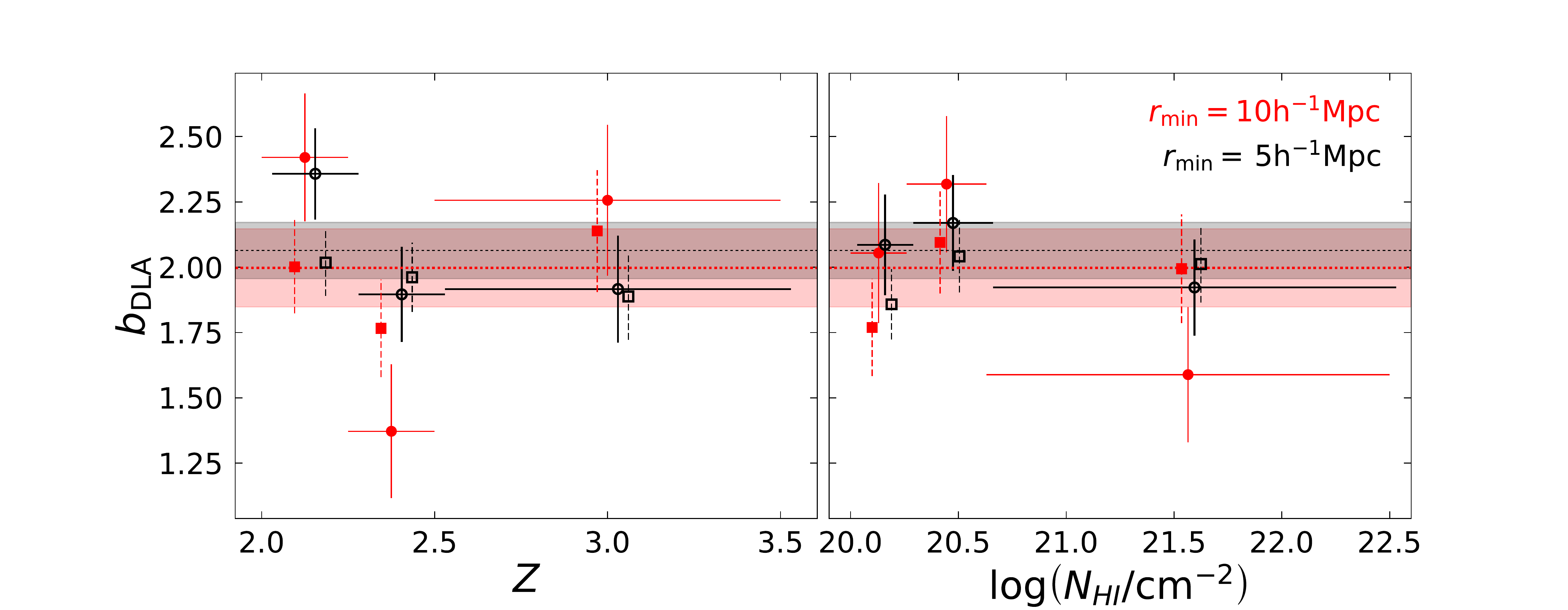}
	\caption{DLA bias versus redshift (left) and
$\log\left(N_{HI}\right)$ (right) obtained from subsamples Z1, Z2, and
Z3, and N1, N2, and N3, respectively
(see table \ref{ta DLA: sample properties}). Black open circles and red closed circles with solid error bars are fit results for $r_{\rm min}=5$ and $10\hmpc$,
respectively. Dotted lines are the result for sample A,
with $1-\sigma$ errors indicated by shaded regions. Squares are equivalent to circles and
triangles, but computed from samples ZC1 to ZC3, and NC1 to NC3, described
in section \ref{subs DLA: bias evolution}. The bins in redshift and
column density are the same for all cases (shown only for the solid
errorbars). Except for red solid circles, points are horizontally shifted to
avoid overlap.}
	\label{fig DLA: dla bias dependence}
\end{figure*}

 The left panel in figure \ref{fig DLA: dla bias dependence} shows the
the DLA bias in three redshift bins, derived from the
cross-correlations of samples Z1, Z2 and Z3 (see table
\ref{ta DLA: sample properties}).
The results are shown with solid errorbars, with horizontal ones
indicating the redshift range of each subsample, for both values of
$r_{\rm min}$, and are also tabulated in table \ref{ta DLA: biases}.
There is no evidence for any redshift evolution of the DLA bias. For
$r_{\rm min}=5\hmpc$, the scatter of the DLA bias in the three redshift
bins is a bit larger than expected, but we believe this is attributable
to statistical noise (using the bootstrap errors the measured scatter
corresponds to a $\sim 2-\sigma$ fluctuation). Results for the larger
sample C1 split into three redshift bins, shown with dotted error
bars, give a smaller scatter; these will be presented in more detail
later in section \ref{subs DLA: bias evolution}.

  The dependence of the DLA bias on column density, obtained from
the subsamples N1, N2 and N3, is shown in the right panel of figure
\ref{fig DLA: dla bias dependence}, with values tabulated in table
\ref{ta DLA: biases}. Again, there is no evidence for any dependence on
$N_{\rm HI}$ for either of the two values of $r_{\rm min}$.

\subsection{Scale dependence of the bias factor} \label{subs DLA: bias r}

  We now test if our measured cross-correlation agrees with the
theoretically expected radial dependence in linear theory of the
$\Lambda$CDM model for $P_L(k)$. If this model is correct there should
be no radial dependence of $b_{\rm DLA}$, except at small scales where
non-linear effects may be important. We repeat the fit of the sample A
cross-correlation to our fiducial model restricted to bins in rings in
the $(r_\parallel , r_\perp )$ plane, defined by
$2^{(i-1)/2}r_{\rm min} < r <  2^{i/2}r_{\rm min}$, with
$i = 0, 1, 2, ..., 8$.

\begin{figure*}
	\centering
	\includegraphics[width=\textwidth]{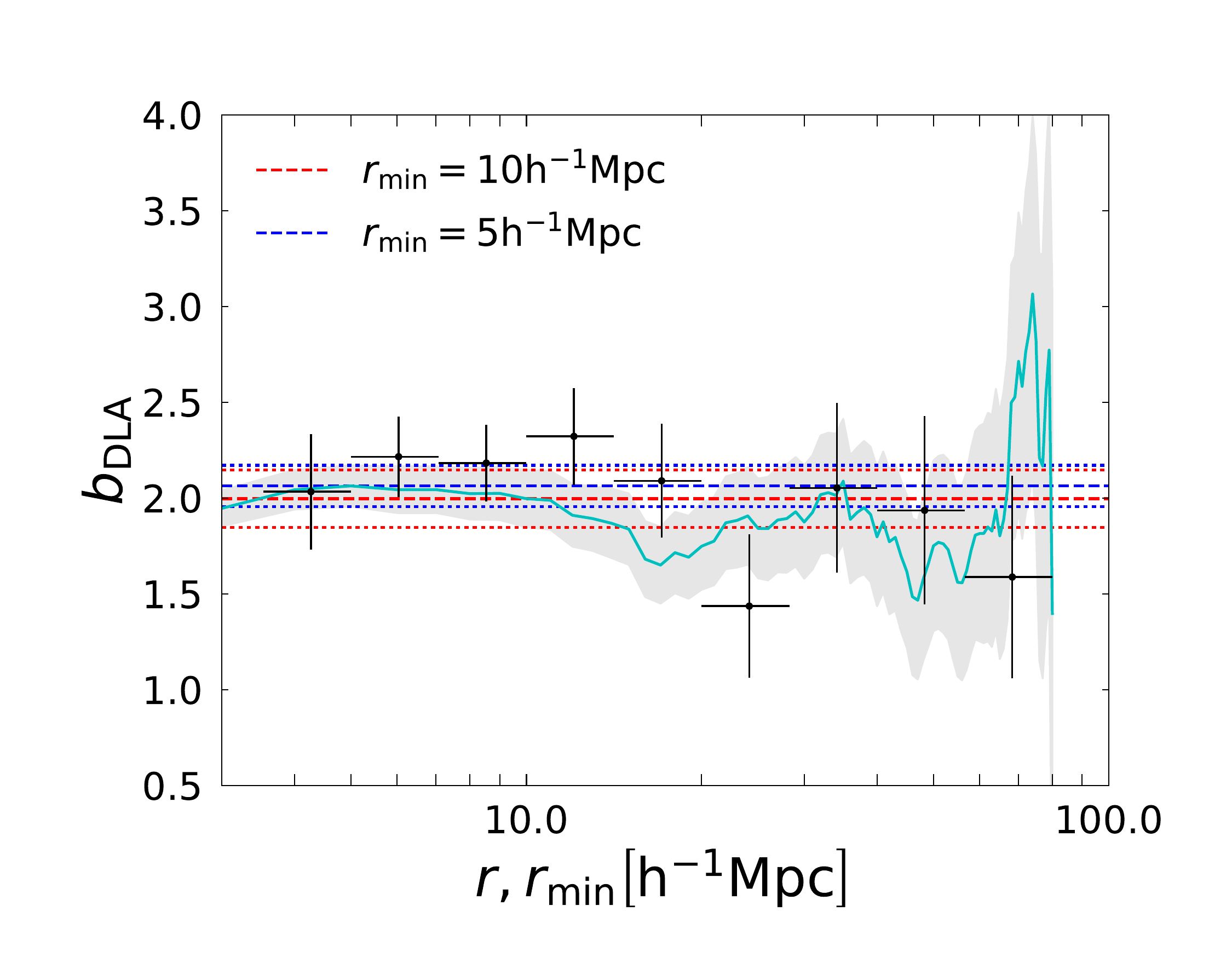}
	\caption{DLA bias versus $r$ obtained by fitting the sample A
cross-correlation in the bins in $r=\sqrt{ r_\parallel^2 + r_\perp^2}$
indicated by the horizontal error bars. Dashed lines show values
obtained by fitting the whole radial range with two different
$r_{\rm min}$, with the $1-\sigma$ error indicated by the dotted lines.
The cyan solid line shows $b_{\rm DLA}$ as a function of $r_{\rm min}$,
the minimum radius of bins included in the fit. The maximum value of $r$
is fixed at $90\hmpc$. Gray area shows the $1\sigma$ confidence levels
around the cyan line.}
	\label{fig DLA: dla bias dependence r}
\end{figure*}

\begin{table*}
	\centering
	\begin{tabular}{llcr}
	\toprule
		$r_{\rm min}$ & $r_{\rm max}$ & $b_{\rm DLA}$ & $\chi^{2}$ (d.o.f.) \\ 
	\midrule
3.54 & 5.00 & $2.03\pm0.30$ & 0.97 (2-1) \\
5.00 & 7.07 & $2.22\pm0.21$ & 20.19 (8-1) \\
7.07 & 10.00 & $2.18\pm0.20$ & 16.75 (24-1) \\
10.00 & 14.14 & $2.32\pm0.25$ & 35.74 (38-1) \\
14.14 & 20.00 & $2.09\pm0.30$ & 83.19 (80-1) \\
20.00 & 28.28 & $1.44\pm0.37$ & 163.14 (154-1) \\
28.28 & 40.00 & $2.05\pm0.44$ & 330.91 (320-1) \\
40.00 & 56.57 & $1.94\pm0.49$ & 640.28 (620-1) \\
56.57 & 80.00 & $1.59\pm0.53$ & 1227.02 (1260-1) \\\bottomrule
	\end{tabular}
	\caption{DLA bias versus $r$ from sample A with the fit
restricted to bins with $r\in\left[r_{\rm min}, r_{\rm max}\right)$, in
units of $\hmpc$.}
	\label{ta DLA: dla bias dependence r}
\end{table*}

 The results of these fits are shown in figure
\ref{fig DLA: dla bias dependence r} and table 
\ref{ta DLA: dla bias dependence r}. 
While there is no clear dependence of the DLA bias on $r$, and most of the values of $\chi^{2}$ are consistent with a good fit for all the rings, we note that the $\chi^{2}$ value for the second ring is particularly bad. The probability of obtaining such a value is about half a percent. If this was our only measurement, then the bad $\chi^{2}$ might indicate that the linear model is starting to fail at these small scales, but see 
\cite{Lochhaas+2016}.
for a more detailed analysis.
However, obtaining such a high $\chi^{2}$ in one out of nine measurements is not as unlikely. This suggests that the linear theory $\Lambda CDM$
model correctly predicts the cross-correlation we have measured, as
expected if DLAs are associated with dark matter halos that trace the
underlying dark matter distribution \citep[e.g.][]{Mo+1996b}.
The cyan line in \ref{fig DLA: dla bias dependence r} shows the result
of cumulative fits to all $r<r_{\rm min}$, with the grey band indicating
the $1-\sigma$ error. This error increases with $r_{\rm min}$ as the
radial range of the fit is reduced.

 We note that at small scales, there is no clear variation of the DLA
bias from linear theory down to the smallest radii we test, as might be
expected from non-linearities. The saturation of absorption lines in
the \lya\ forest naturally acts as a mask of the contribution from
highly overdense regions to cross-correlations, making linear theory
predictions surprisingly accurate down to rather small scales. We
therefore consider that our results for the DLA bias with
$r_{\rm min} > 10 \hmpc$ are not significantly affected by
non-linearities in the cross-correlation. The lack
of any clearly visible spreading of contours in $r_\parallel$ at small
scales in figure \ref{fig DLA: cross_correlation contour} also shows
that the combination of intrinsic velocity dispersions and redshift
errors in our sample A of DLAs is small.

%
\section{Comparison with previous results and model dependence of the DLA bias} \label{sec DLA: comparison DR9}

  We now analyse in detail the model dependence of our result on the
mean DLA bias. To facilitate the comparison with the previous result
of FR12, our reference result in this section will be for sample C2
and $r_{\rm min}= 5\hmpc$, which is $b_{\rm DLA}= 1.99\pm 0.09$ from
table \ref{ta DLA: biases}. FR12 used an equivalent sample for DR9,
and the same value of $r_{\rm min}$, and obtained
$b_{\rm DLA}= 2.17 \pm 0.20$.
However, the difference with our result does not just arise from using
DR12 instead of DR9, but from the following differences in the analysis
and the model that is fitted to the data:
\begin{enumerate}
	\item The bias fitting method.
	\item The cross-correlation estimator and covariance matrix.
	\item Correction for the continuum fitting distortions.
	\item Larger DR12 dataset versus DR9.
	\item Cosmological model.
	\item \lya{} forest bias parameters.
	\item New additions to the cross-correlation model.
\end{enumerate}
In the rest of this section, we start with the original result of FR12
and change these factors one by one to see how each of them affects the
result of $b_{\rm DLA}$. The last three points also account for the main
model dependence of our result, discussed in subsections
\ref{subs DLA: cosmo} to \ref{subs DLA:model dependence}.
To help the reader track all the effects and changes caused on
$b_{\rm DLA}$, a list is provided in table \ref{ta DLA: corrections}.

\subsection{Bias fitting method}\label{subs DLA: fitter}

  The fitting of the model to the data was done in FR12 with an MCMC
code written especially for that paper. We have used in this paper the
BAOFIT code \citep{Kirkby+2013}, which computes errors from the
second derivatives of the $\chi^2$ function computed from the covariance
matrix that is provided. There may therefore be slight differences in
the results obtained with the two codes. To test this difference, we
have run BAOFIT to fit the $b_{\rm DLA}$ parameter with exactly the same
values of the cross-correlation over the same bins, and the same
covariance matrix that was computed in FR12 from the DR9 data.

  The continuum fitting method of FR12 was different than the one used
here, and a correction of the continuum fitting effects called the {\it Mean Transmission Correction} (hereafter, MTC) was applied there to
the fitted model, which was quite different from our distortion matrix
correction described in section \ref{subs DLA: Distortion matrix}. To
take out differences in the fitted model, we compare to the FR12 result
for their NOCOR case, in which no MTC correction was applied to the
model, and the value obtained was $b_{\rm DLA}= 2.00 \pm 0.19$. We use
exactly the same cosmological model and \lya{} forest bias parameters as
were used in FR12, and we eliminate the factors $G(\bf k)$ and
$S(k_{\parallel})$ in equation \ref{eq DLA: PS model} and our
corrections for the distortion matrix and the presence of HCDs, to fit
to exactly the same cross-correlation model as in FR12. Our result is
$b_{\rm DLA}= 2.01 \pm 0.17$.

 We therefore conclude that the main effect of the different fitting
method is that the errorbars from BAOFIT using the covariance matrix
are $\sim 10\%$ smaller than those from the MCMC code used in FR12. 

\subsection{The cross-correlation estimator and covariance matrix }
\label{subs DLA: estimator/covmat}

 We now use our own method to determine the continuum of the observed
spectra and the values of the \lya{} transmission, and to estimate the
cross-correlation and covariance matrix, using only sample C2 limited to
the DR9 dataset, i.e., the same data used by FR12. Our \lya{} forest data
also includes the masking and correction for DLAs in the catalogue of
\cite{Noterdaeme+2014}, which were not included in FR12. We fit
$b_{\rm DLA}$ exactly as before, using the same cosmological model and
\lya{} forest bias parameters as FR12, and not including any of the
corrections that were not included in FR12. The result we find is
$b_{\rm DLA}=1.94\pm 0.15$.

 We conclude that the difference due to the estimator and covariance
matrix (comparing again to the NOCOR case of FR12) is that the DLA bias
we obtain is $\sim0.07$ lower, or reduced by 3.5\%, and the error is
20\% smaller, compared to FR12. This must be caused by the different
way of fitting the continuum to obtain the \lya{} transmission, 
the correction of detected DLAs in the data, and the different
covariance matrix we use. We note that of the 20\% reduction in the
error, 10\% is due to the different fitting code as found above.

\subsection{Correction for the continuum fitting distortion }
\label{subs DLA: distortion}

 Next, we include the correction for the continuum fitting distortion.
In FR12, the inclusion of their MTC correction to the fitted model
modified the derived bias from $b_{\rm DLA}=2.00\pm 0.19$ for their
NOCOR case, to $b_{\rm DLA}=2.17\pm 0.20$, which was the fiducial or
main result in that paper. In our case, using the BAOFIT code and our
own estimate of the cross-correlation and covariance matrix, including
the distortion matrix method introduced by \citet[][see section
\ref{sec DLA: proj} for a more detailed explanation]{Bautista+2017}
raises the derived bias from $b_{\rm DLA} = 1.94 \pm 0.15$ to
$b_{\rm DLA} = 2.14 \pm 0.16$.

 We therefore conclude that the two different corrections for the
distortions introduced by continuum fitting are very similar. The
difference in the derived bias factor when we combine the effects of the
fitting method, the estimation of the cross-correlation and covariance
matrix, and the continuum fitting distortion corrections, is reduced to
only $0.03$, and our error based on the covariance matrix is 20\%
smaller than in FR12 for the reasons discussed in the previous
subsections. The fact that two completely independent methods to correct
continuum fitting distortions are in good agreement increases our
confidence in the accuracy of this correction.

\subsection{Larger DR12 dataset versus DR9}
{\label{subs DLA: sample}

 We now change the dataset from DR9 to DR12, using as before sample C2
with $r_{\rm min} = 5\hmpc$, and our method to evaluate the
cross-correlation and covariance matrix, and the distortion matrix to
correct for the continuum fitting effect. The result is that the DLA
bias decreases from $b_{\rm DLA}= 2.14 \pm 0.16$ for DR9, to
$b_{\rm DLA}= 2.02 \pm 0.09$ for DR12. This change between the two data
samples is consistent with the expected statistical error, and the
decrease in the error bar is as expected from the increase of the
sample size.

  The increased size of the sample from DR9 to DR12 has therefore caused
a decrease of the measured DLA bias of $0.75$ times the error we infer
for DR9. However, we shall now see that systematic differences in the
model of FR12 and our own cause larger changes on $b_{\rm DLA}$.

\subsection{Cosmological model }\label{subs DLA: cosmo}

 Next, we repeat the fit to $b_{\rm DLA}$ for the C2 sample of DR12,
applying as before our distortion matrix, and we change the cosmological
model from the one used in FR12 based on WMAP results (with parameters
$\Omega_m=0.281$ and $\sigma_8=0.8$) to the Planck model we use here,
with $\Omega_m=0.3156$ and $\sigma_8=0.831$. The bias changes from
$b_{\rm DLA}= 2.02 \pm 0.09$ to $b_{\rm DLA}= 1.80 \pm 0.08$.

 There are two main reasons for this change. First, the normalization of
the power spectrum at the reference redshift $z_{\rm ref}=2.3$, which is
close to the mean redshift where the DLA-\lya{} cross-correlation is
measured, is proportional to $b_{\rm DLA}$ times the square of the rms
density fluctuation at $z_{\rm ref}$. This fluctuation is usually
expressed in terms of its average over a sphere of radius $8\hmpc$,
which we find to be $\sigma_8\left(z_{\rm ref}\right)=0.3120$ for the FR12 model,
and $\sigma_8\left(z_{\rm ref}\right)=0.3161$ in our Planck model. This therefore
implies a reduction of $b_{\rm DLA}$ by a factor
$(0.3120/0.3161)^2=0.974$, if we neglect the small change in
$\beta_{\rm DLA}$ corresponding to a change in $b_{\rm DLA}$. We note
that the scale of a sphere of $8 \hmpc$ radius is close to the effective
scale at which our cross-correlation is measured, so apart from the
normalization parameter $\sigma_8\left(z_{\rm ref}\right)$, there is little
variation of $b_{\rm DLA}$ due to the small change in the shape of the
power spectrum between the two models. For instance, a $2\sigma$ change in $n_{s}$ does not significantly change the value of the recovered bias.

  The second reason is the change in the angular diameter distance and
Hubble constant. Our measurements of the cross-correlation are made at
known angular and redshift separations, whereas the model correlation
function is predicted in comoving coordinates in units of $\hmpc$. The
ratio of the quantity $H_0 D_A\left(z_{\rm ref}\right)$ in the model used in this
paper and the FR12 model is $0.9690$, and the ratio of the quantity
$H_0/H\left(z_{\rm ref}\right)$ for our model and the FR12 model is $0.9484$. We take
an average of these two scaling factors, $\sim 0.96$, as the
characteristic ratio by which the comoving scale that is computed from
observed angular and redshift separations changes between the two
cosmological models. The model $\Lambda CDM$ cross-correlation varies
approximately as $\xi \sim r^{-2}$ over the range of scales in which
our measurement is most significant, so this implies an approximate
reduction in the inferred $b_{\rm DLA}$ by a factor $\sim 0.92$.
Combining this with the previous reduction factor from
$\sigma_8(z_{\rm ref})$, we see how a total reduction of the inferred
$b_{\rm DLA}$ by $\sim$ 10\% due to the change of the cosmological model
is explained.

\subsection{\lya{} forest bias parameters }
\label{subs DLA: lya}

 Apart from the cosmological model, our result on $b_{\rm DLA}$ is also
strongly affected by the \lya{} forest bias parameters. The bias
parameters used in FR12 were $\beta_{\rm Ly\alpha}=1$ and
$b_{\rm Ly\alpha}\left(1+\beta_{\rm Ly\alpha}\right) = -0.336$ at a reference
redshift $z_{\rm ref}=2.25$, taken from \cite{Slosar+2011}. This needs
to be transformed to the reference redshift we use of $z_{\rm ref}=2.3$,
using the assumed evolution of the \lya{} forest bias of
$b_{\rm Ly\alpha} \propto (1+z)^{2.9}$ in all the papers that have
measured the \lya{} autocorrelation. The result is
$b_{\rm Ly\alpha}(1+\beta_{\rm Ly\alpha}) = -0.351$.

  These values were updated first by the analysis of
\cite{Blomqvist+2015}, who fitted the DR11 \lya{} autocorrelation applying
the linear theory model only to scales $r> 40 \hmpc$. Their result was
$b_{\rm Ly\alpha}\left(1+\beta_{\rm Ly\alpha}\right) = -0.374$, and
$\beta_{\rm Ly\alpha}=1.39$, at $z_{\rm ref}=2.3$. Using these \lya{} bias
parameters, the DLA bias changes only from $b_{\rm DLA}=1.80 \pm 0.08$
to $b_{\rm DLA}=1.82 \pm 0.08$. However, the more recent analysis of
DR12 by \cite{Bautista+2017} gives a substantially different result,
when fitting all the data down to $r> 10 \hmpc$ to the same model as
before (model labelled Ly$\alpha$ in their table 5):
$b_{\rm Ly\alpha}\left(1+\beta_{\rm Ly\alpha}\right) = -0.326$, and
$\beta_{\rm Ly\alpha}=1.246$, which makes our result for the DLA bias
increase to $b_{\rm DLA}=2.05\pm 0.09$. The reason for this change from
\cite{Blomqvist+2015} is that the \lya{} autocorrelation data prefers a
lower \lya{} bias factor at smaller scales, and in fact
\cite{Bautista+2017} noted that this simple model does not provide a
good fit to the whole radial range.

\begin{figure*}
	\centering
	\includegraphics[width=\textwidth]{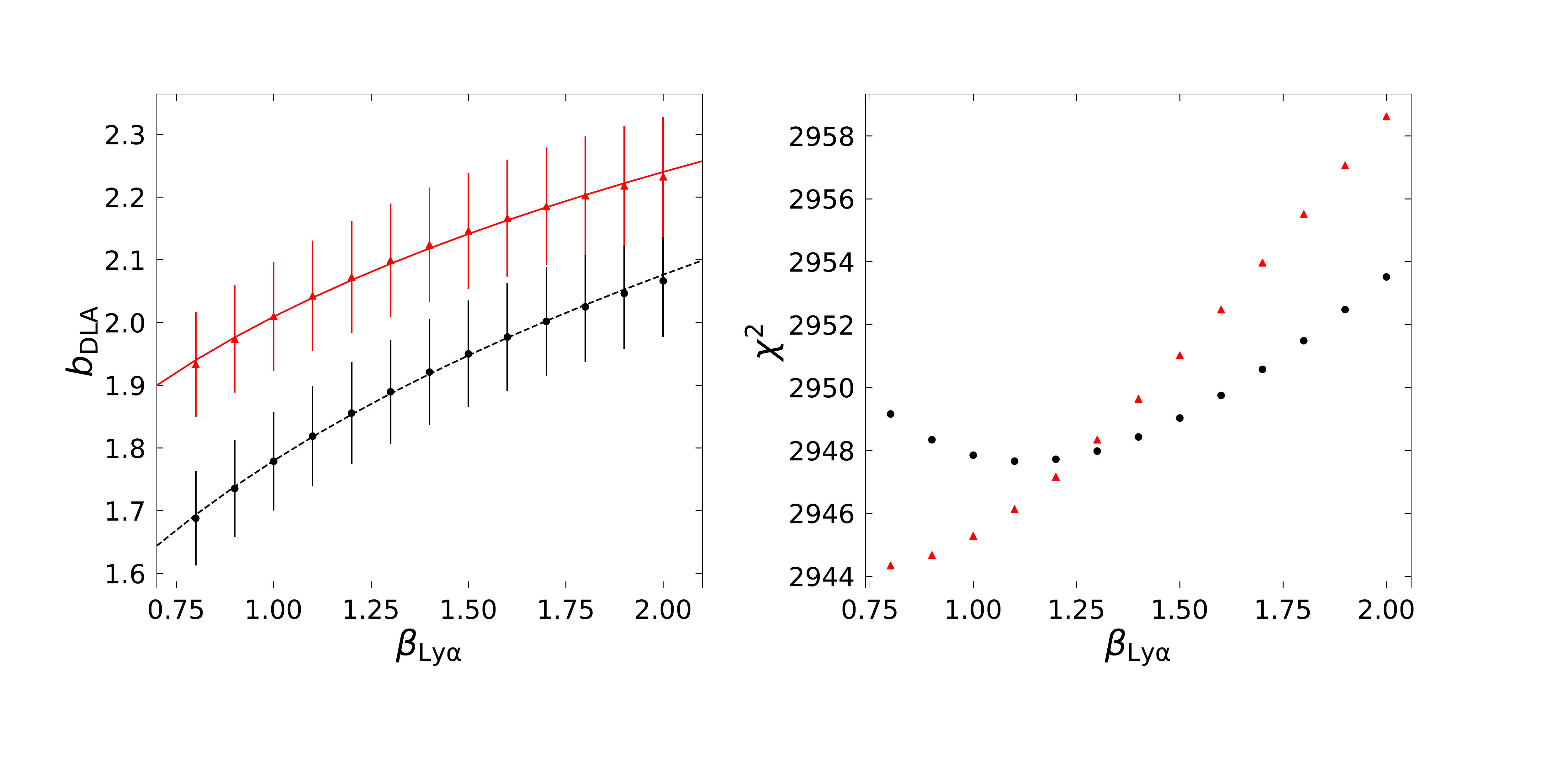}
	\caption{{\it Left}: Inferred $b_{\rm DLA}$ as a function of
$\beta_{\rm Ly\alpha}$ when keeping the fixed value
$b_{\rm Ly\alpha}(1+\beta_{\rm Ly\alpha})=-0.325$. Red triangles do not
include the HCD correction, and black circles include it. Also shown
are power-law fits $b_{\rm DLA} \propto \beta_{\rm Ly\alpha}^{\gamma_1}$,
with $\gamma_1=0.21$ for no HCD correction (solid red line) and
$\gamma_1=0.23$ with the correction (dashed black line). {\it Right}:
Values of $\chi^{2}$ when fitting $b_{\rm DLA}$ for different
$\beta_{\rm Ly\alpha}$, with the HCD correction (black circles) and
without (red triangles).}
	\label{fig DLA: dla vs beta}
\end{figure*}

  The dependence of our result on the \lya{} forest bias parameters can
be understood by noting that the amplitude of the cross-correlation
model is proportional to $\sigma_8^2\left(z_{\rm ref}\right)\, b_{\rm Ly\alpha}\,
b_{\rm DLA}$. Only this product can be inferred from the
cross-correlation measurement. However, the angular dependence of the
redshift distortion factors introduces a more complex dependence on
$\beta_{\rm Ly\alpha}$ and $\beta_{\rm DLA}$. We show in figure
\ref{fig DLA: dla vs beta} the inferred $b_{\rm DLA}$ as a function of
$\beta_{\rm Ly\alpha}$, when keeping
$b_{\rm Ly\alpha}\left(1+\beta_{\rm Ly\alpha}\right)$ fixed, as the red circles
(black circles include the HCD correction and are discussed below).
Error bars are our statistical errors from the cross-correlation
measurement. Following FR12, we fit a power-law dependence
$b_{\rm DLA}\propto \beta_{\rm Ly\alpha}^{\gamma}$, finding
$\gamma=0.21$ over the range of interest shown in figure
\ref{fig DLA: dla vs beta}.

 The variation of $b_{\rm DLA}$ is nearly proportional to
$b_{\rm Ly\alpha}^{-1}$ at fixed $\beta_{\rm Ly\alpha}$, except for the
fact that $\beta_{\rm DLA}\propto b_{\rm DLA}^{-1}$, implying that
$b_{\rm DLA}$ increases a bit faster than expected with decreasing
$b_{\rm Ly\alpha}$ because of the need to compensate for a smaller
redshift distortion factor.

\subsection{New additions to the cross-correlation model}
\label{subs DLA:model dependence}

 Our model incorporates improvements that were not present in FR12: the
correction for binning of the cross-correlation and the wavelength PSF,
and the HCD correction. The size of the bins in $r_{\parallel}$ and
$r_{\perp}$ of $2\hmpc$ is corrected for by multiplying by the function
$G\left({\bf k}\right)$ in Fourier space (see section
\ref{sec DLA: Model} and equation \ref{eq DLA: PS model}). This
increases our last value $b_{\rm DLA}=2.05\pm 0.09$ for the linear
\lya{} model of \cite{Bautista+2017} to $b_{\rm DLA}=2.06\pm 0.09$.
The wavelength PSF includes the BOSS spectrograph resolution and the
rebinning of the BOSS spectral pixels into analysis pixels that are three
times wider, as discussed in section \ref{sec DLA: Model}, and is
corrected by multiplying by the function $S\left(k_{\parallel}\right)$. This further
increases our result to $b_{\rm DLA}=2.08\pm 0.09$.

 A more important impact on the DLA bias is caused by the HCD
correction, introduced by \cite{Bautista+2017} and discussed in section
\ref{sec DLA: Model}. The black crosses in figure
\ref{fig DLA: dla vs beta} show the inferred $b_{\rm DLA}$ as a function
of $\beta_\alpha$ when the HCD correction is included, and the power-law
fit for this case, with $\gamma=0.23$, is shown as the dashed black
line.  At a fixed value of $\beta_\alpha$, including the HCD correction
causes a reduction of $\sim$ 10\% on $b_{\rm DLA}$. However, this
correction must be included self-consistently with the parameters fitted
to the \lya{} autocorrelation. We therefore change to the final values of
the \lya{} forest bias factors we use for our fiducial result of the DLA
bias in this paper, those in table 3 of \cite{Bautista+2017}: 
$b_{\rm Ly\alpha}\left(1+\beta_{\rm Ly\alpha}\right) = -0.326$ and
$\beta_{\rm Ly\alpha}=1.663$, which were obtained by including the HCD
correction in the fitted model, but also an additional correction due to
metal lines. For these values, we find $b_{\rm DLA}=2.18\pm 0.10$ when
not including the HCD correction in the cross-correlation fit, and
$b_{\rm DLA}=1.99\pm 0.09$ when including it (corresponding to the red
and black curves in figure \ref{fig DLA: dla vs beta}, respectively, at
$\beta_{\rm Ly\alpha}=1.663$). We have generally not included the
correction for metal lines in this paper, because their effect is not
detected in our DLA-\lya{} cross-correlation. However, we find that
including the same metal-line correction in our analysis increases the
DLA bias to $b_{\rm DLA}=2.01\pm 0.09$, and worsens the $\chi^2$ value.

  The dependence of $b_{\rm DLA}$ on the HCD correction is therefore
substantially smaller than 10\% when we use self-consistently the values
of the \lya{} bias factors that fit the \lya{} autocorrelation. The reason
why $\beta_{\rm Ly\alpha}$ needs to increase when including the HCD correction
is that the latter adds to the cross-correlation model a function that
is elongated in the parallel direction, accounting for the Voigt
profiles with damped wings of HCD absorbers. This needs to be
compensated by an increased Kaiser effect in the linear model, causing a
tangential elongation. The change in $b_{\rm DLA}$ from the model with
$\beta_\alpha=1.246$ and no HCD correction, to the model with
$\beta_\alpha=1.663$ with the HCD correction, is less than 5\% (from
$2.08$ to $1.99$), and reflects the true impact of the HCD correction. 

 Finally, the right panel of figure \ref{fig DLA: dla vs beta} shows the
$\chi^2$ value of our fit as a function of $\beta_{\rm Ly\alpha}$,
with no HCD correction (red triangles) and including it (black circles).
It is interesting that the best fit value for no HCD correction,
$\beta_{\rm Ly\alpha}=1.1\pm 0.3$, is lower than that obtained by
\cite{Bautista+2017} (although only at the $1.5-\sigma$ level with the
HCD correction), and that the HCD correction worsens our fit by
$\Delta \chi^2\simeq 4$. This is probably an indication that the HCD
correction is not a sufficiently good model of the impact of HCD
absorption wings in the \lya{} spectra.

\begin{table*}
	\centering
	\begin{tabular}{ccc}
		\toprule
		Introduced correction & $b_{\rm DLA}$ & $\chi^2 (d.o.f)$ \\
		\midrule
		Original FR12, NOCOR & $2.00\pm 0.19$ &  \\
		Use BAOFIT for model fitting & $2.01\pm 0.17$ &  \\
		Use our $\xi_{A}$, $C_{AB}$ & $1.94\pm 0.15$ &  \\
		Distortion matrix correction & $2.14\pm 0.16$ &  \\
		(Original FR12, fiducial) & $2.17\pm 0.20$ &  \\
		From DR9 to DR12 & $2.02\pm 0.09$ &  \\
		Change to Planck-2016 cosmological model & $1.80\pm 0.08$ &  \\
		\cite{Bautista+2017} \lya\ bias factors & $2.05\pm 0.09$
		 & 2,954.41 (2,864-1)\\
		Smoothing correction G & $2.06\pm0.09$ & 2,952.14 (2,864-1)\\
		Smoothing corrections $G\cdot S$ & $2.08\pm 0.09$
		 & 2,947.73 (2,864-1)\\
		HCD correction with final \lya\ bias factors
                 & $1.99\pm 0.09$ & 2,950.26 (2,864-1)\\
		HCD and metal corrections
                 & $2.01\pm 0.09$ & 2,954.12 (2,864-1)\\
		\bottomrule
	\end{tabular}
	\caption{Summary of all the effects contributing to the difference
from the result of FR12 and the final result obtained here for
$b_{\rm DLA}$. Errors are obtained from the covariance matrix. The
intermediate result after applying the distortion matrix correction
needs to be compared to the fiducial result of FR12,
$b_{\rm DLA}=2.17\pm 0.20$, to see that our methods produce very similar
results when applied to the same data with the same fitting model. All
our results are for the C2 sample. }
	\label{ta DLA: corrections}
\end{table*}

 To summarize all the differences from FR12 and model dependences
discussed in this section, table \ref{ta DLA: corrections} lists all
the changes of $b_{\rm DLA}$ caused by each of the effects we have
discussed.

%
\section{Discussion}  \label{sec DLA: Discussion}

\subsection{Systematic errors: Cross-correlation modelling}
\label{subs DLA: systematics}

 So far, all of the errors we have quoted for $b_{\rm DLA}$ include only
statistical errors of the cross-correlation measurement, computed either
from our covariance matrix or the bootstrap analysis. We now discuss
systematic errors, which arise from two sources: uncertainties in the
model of the cross-correlation to be used in the fit, and impurity of
the DLA sample. We discuss first the modelling uncertainties.

 There are several possible sources of systematic error of $b_{\rm DLA}$
in our modelling procedure: the continuum fitting correction, the assumed
cosmological model, the use of linear theory, the \lya{} bias factors,
and the HCD correction. We believe our continuum fitting correction is
accurate, in view of the good agreement of two independent methods of
applying this correction from FR12 and the distortion matrix procedure
used here (see section \ref{subs DLA: distortion}), and the tests that
have been made with mocks \citep{Bautista+2017}. While we have shown
that there is a high sensitivity to the cosmological model (a 10\%
variation of $b_{\rm DLA}$ is caused by the update from the WMAP model
of FR12 to the Planck model we use), this does not cause a large
systematic if we believe that the results of \cite{Planck2015} are
accurate. As we shall see below, we are particularly interested in
systematics that might lower our inferred value of $b_{\rm DLA}$, to
bring it in closer agreement with expectations from cosmological
simulations of galaxy formation, and this can only occur by further
increasing $\sigma_8(z_{\rm ref})$ or $\Omega_m$ in the cosmological
model. Linear theory seems well justified from the constant value of
$b_{\rm DLA}$ with scale (figure \ref{fig DLA: dla bias dependence r})
and the large value of $r_{\rm min}$ we are using, although precise
predictions of non-linearities in the DLA-\lya{} cross-correlation from
cosmological simulations would be highly desirable to test this.

  We believe the more important sources of systematics are in the
uncertainties in the \lya{} bias factors and the HCD correction
determined from the \lya{} autocorrelation. If we use the two models
fitted to the \lya{} autocorrelation in \cite{Bautista+2017}, the
'Ly$\alpha$' one in their table 5 without HCD correction and
$\beta_{\rm Ly\alpha}=1.246$, and the full model in their table 3 with
HCD correction and $\beta_{\rm Ly\alpha}=1.663$, the difference
of $\sim$ 5\% in the implied $b_{\rm DLA}$ between the two models is a
good estimate of our systematic error, which is comparable to our
statistical error of $b_{\rm DLA}$. The model including the HCD
correction gives the lowest value of $b_{\rm DLA}$, and we
conservatively take it as our final result to compare with predictions
from galaxy formation simulations. The statistical errors in the \lya{}
bias factors of the models of \cite{Bautista+2017} are negligibly small
for our purpose.

  We note that even though the HCD correction is not a very
accurate representation of the true effect of HCDs, since it
does not take into account the precise Voigt profile shape of the
absorbers, we believe the important thing is that the same model that
provides a good fit to the measured \lya{} autocorrelation is used to
model the DLA-\lya{} cross-correlation.
 
\subsection{Systematic errors: Sample Purity and Selection}
\label{subs DLA: purity samples}

 We now discuss the purity and selection effects of our samples A, C1 and C2.
DLAs are detected using the automatic algorithm described in
\cite{Noterdaeme+2009,Noterdaeme+2012b}, which searches for regions of
strong absorption that are consistent with a damped \lya{} absorption
line plus the random absorption by the \lya{} forest. We therefore expect
that some fraction of these candidate DLAs in the catalogue are
not real DLAs, and are likely to be instead regions of strong absorption
over a sufficiently broad velocity interval to look approximately like a
damped profile, but with a real HI column density much less than
$10^{20}\, {\rm cm}^{-2}$. These false detections should increase at low
$N_{\rm HI}$ and low signal-to-noise. For very low signal-to-noise some
false DLAs may not correspond to any absorber but be mostly
caused by noise, but these cases should be rare with our imposed
cut of $CNR > 3$.

 The fact that we observe no variation of $b_{\rm DLA}$ with column
density suggests that the effect of these contaminants on $b_{\rm DLA}$
is not large. Either the impurity is too small to affect our result, or
the false DLAs are regions of absorption with a bias that is close to
that of DLAs. An additional argument against a large level of
impurity of our sample is the result of \cite{Mas-Ribas+2017} on the
dependence of the mean equivalent width of high-ionization lines of DLAs
on $N_{\rm HI}$ (see their figure 11), which varies only by 20\% over
the available range of column densities. Moreover, this small variation
is not necessarily due to a variation of the impurity level, but can be
caused by a real physical effect.

 A more detailed study requires predicting the purity of our catalogue
using \lya{} forest mock spectra. This is not a simple calculation,
because the mock spectra must have the correct distribution of HCDs with
broad absorption features that can mimic DLAs (which our current mocks
are not designed to reproduce), and DLAs must be inserted in the mock
spectra with the correct cross-correlation, so we leave this for future
studies.

 Different issues arise with DLAs detected blueward of the \lyb{} quasar
emission line, where the \lyb{} forest is superposed with \lya{}
absorption and the possibilities of confusion increase. Often, a DLA may
be detected in part because there is a \lyb{} line from an absorber that
has most of the column density, resulting in an incorrect assigned
redshift. In the absence of any real \lya{} absorber, these absorbers
with incorrect redshifts should contribute a zero cross-correlation to
our measurement, decreasing the fitted value of $b_{\rm DLA}$ by a
fractional amount equal to the fraction of these systems. Our sixth
cut, applied to define sample A, eliminates these misidentified objects,
which are therefore present only in our samples C1 and C2
(see section \ref{sec DLA: Sample data}; sample C1 is the largest, due to
not including cuts 4 and 5 that eliminate DLAs too close to the \ovi{} and
\lya{} quasar emission lines).

 For $r_{\rm min}=5\hmpc$, we measure $b_{\rm DLA} = 2.06 \pm
0.11$ for sample A, $b_{\rm DLA}=1.97\pm 0.08$ for sample $C1$, and
$b_{d} = 1.99\pm 0.09$ for sample C2. These errors are obtained from
our covariance matrix, and are roughly proportional to the inverse
square root of the number of DLAs in each sample. The variation in the
bias of the DLA samples is hardly significant, especially if the
larger bootstrap errors are considered (see table \ref{ta DLA: biases}),
although they go in the expected direction of a decreased $b_{\rm DLA}$
in samples with an expected higher impurity. This suggests that
DLAs removed by cuts 4 to 6 are primarily DLAs or other
absorption systems with a similar bias factor.

\begin{figure*}
	\centering
	\includegraphics[width=0.7\textwidth]{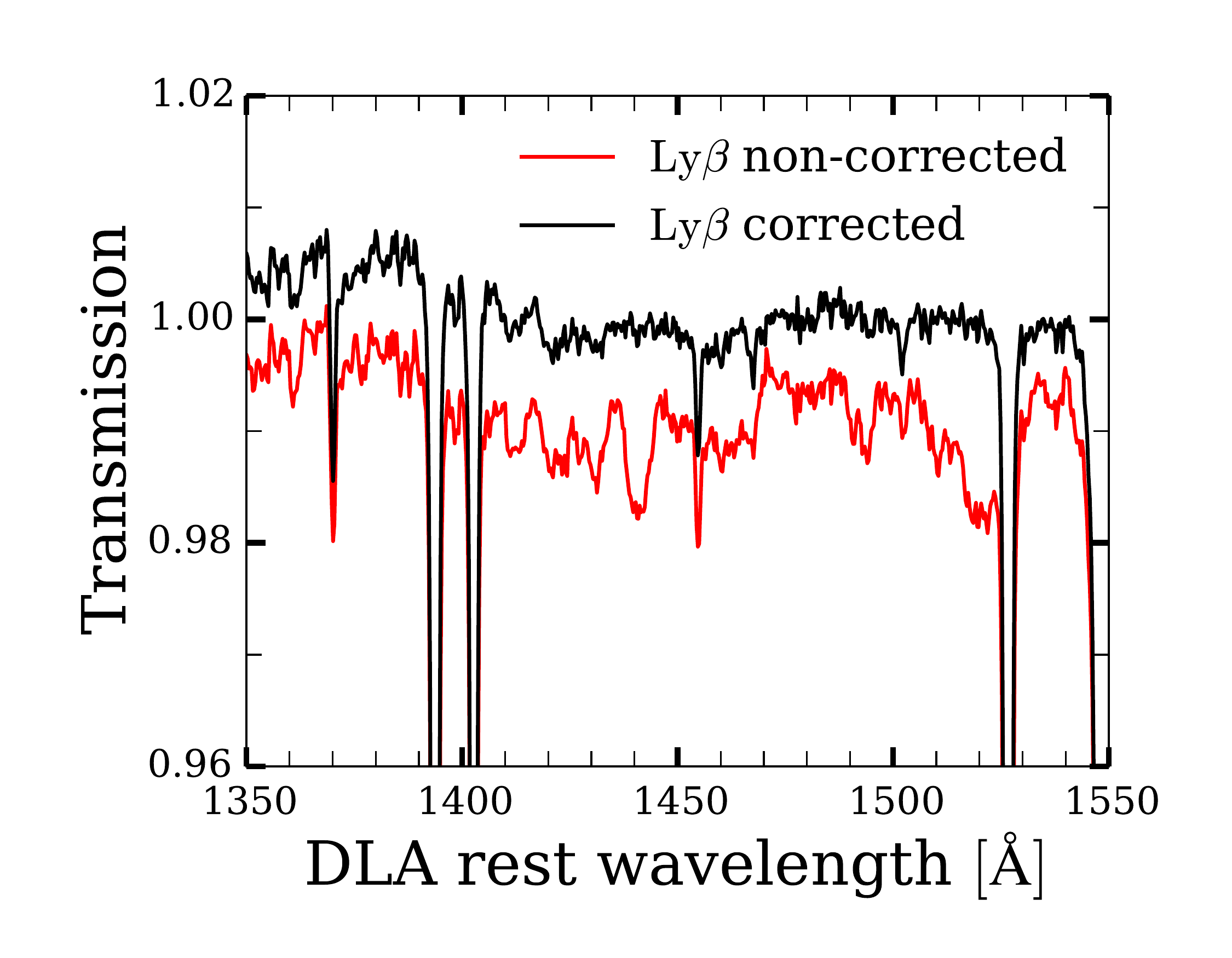}
	\caption{DLA transmission spectrum of the total sample of
\protect\cite{Mas-Ribas+2017} (black line), and the same transmission
spectrum for the larger sample including DLAs in the \lyb{} forest
region, which may include \lyb{} absorbtion lines misidentified as \lya{}
lines of DLAs (thick, red line). The absorption feature at
$\sim 1,441\angs$ suggests that these misidentified DLAs are
$\sim$ 1\% of the sample.}
	\label{fig DLA: lyb contamination}
\end{figure*}

 The level of impurity due to \lyb{} absorption lines confused by \lya{}
n samples C1 and C2 can also be estimated by stacking the absorption
spectra. If a \lyb{} absorption line is incorrectly attributed to \lya{}
absorption at $\lambda_\alpha=1216\, {\rm \AA}$, then the true \lya{}
absorption will appear at $32/27 \lambda\alpha=1441\, {\rm \AA}$.
Figure \ref{fig DLA: lyb contamination}) shows the stacked spectrum
obtained with the technique of \cite{Mas-Ribas+2017} for what was
designated as ``total sample'' by these authors, which excluded DLAs
found in the \lyb{} forest region, as the black line. The red line is
the stacked spectrum of the larger sample including DLAs in the \lyb{}
forest region, and shows the expected absorption of misidentified
absorbers at a level of $\sim $ 1\% . We take this as an upper limit to
the fraction of systems in our C1 and C2 samples that are \lyb{} lines
wrongly identified as \lya{} lines of DLAs, because these \lyb{} lines
are more likely to be identified when they have superposed true \lya{}
absorption. This indicates that this contamination of the sample is
very small and not significant compared to our statistical errors.

 In general, the inclusion of any false absorbers in our catalogue arising
purely from noise or from misidentified redshifts can only decrease our
measured $b_{\rm DLA}$, because the false absorbers have a null average
contribution to the cross-correlation. The presence of HCD absorbers
misidentified as DLAs may increase our measured bias only if the HCD
bias is higher than that of DLAs. There is, however, a systematic
arising from a selection effect that may increase the measured
$b_{\rm DLA}$, already mentioned in FR12: if DLAs are more likely to be
detected when the \lya{} forest that is superposed with their damped
wings is weaker than average, then this would preferentially select DLAs
surrounded by high-density large-scale regions, over those in low
density regions. The reason is that the DLA-\lya{} cross-correlation is
negative along the line-of-sight at $\left| r_\parallel \right| \gtrsim 20 \hmpc$
owing to redshift space distortions, implying weaker \lya{} forest
absorption over the damped wings of DLAs in more overdense regions.
This is a selection effect that can only be properly corrected with the
use of adequate mock spectra with DLAs inserted with the correct
cross-correlation with the \lya{} forest. Again, we believe this
correction is unlikely to be large because of the absence of dependence
of $b_{\rm DLA}$ on $N_{\rm HI}$, but future studies will need to better
address this question.

\subsection{Evolution of the bias factor} \label{subs DLA: bias evolution}

 The lack of a significant dependence of $b_{\rm DLA}$ on redshift and
column density was shown for sample A in figure
\ref{fig DLA: dla bias dependence}, although a large scatter was noticed
for the redshift dependence for the case with $r_{\rm min}=10\hmpc$,
with a lower bias for the middle redshift than the low and high redshift
ones by $\sim 2.5\sigma$. To explore if this scatter might indicate
something other than noise, we repeat the measurement using sample C1,
taking into account that decreased purity is unlikely to be very
important as argued in section \ref{subs DLA: purity samples}.

\begin{table}
	\centering
	\begin{tabular}{cr}
		\toprule
		Name & Number of DLAs \\
		\midrule
		ZC1	& 6,319\\
		ZC2 	& 6,664\\
		ZC3	& 10,359\smallskip\\
		NC1	& 8,613\\
		NC2	& 7,788\\
		NC3	& 6,941\\
		\bottomrule
	\end{tabular}
	\caption{Number of DLAs in the subsamples ZC1 to ZC3 and NC1 to
NC3, drawn from sample C1.}
	\label{ta DLA: sample properties extended}
\end{table}

  We define six new subsamples by dividing the C1 sample into the same
three redshift and column density bins as in table
\ref{ta DLA: sample properties} for sample A. The number of systems in the
new subsamples are shown in table \ref{ta DLA: sample properties extended},
nearly doubling those from sample A.
Results are shown in figure \ref{fig DLA: dla bias dependence}, where
squares with dashed error bars show the bias
values obtained with the new subsamples, and the circles with
solid error bars show the previous results from sample A. The normal
scatter for sample C1 suggests that the anomalously high scatter
in sample A is only due to statistical noise.

The fact that no change of the bias factor (within 10\%) is seen between redshift 2 to 3 suggests that the characteristic host halo mass is decreasing with redshift. The independence with column density also suggests that the mean NHI radial profile is similar in host halos of different masses.

\subsection{Implications on the distribution of DLA host halo masses}
\label{subs DLA: halo mass}

 The bias factor of dark matter halos as a function of their mass is
robustly predicted in analytic models and numerical simulations
\citep[see e.g.][]{Sheth+1999, Tinker+2010}, and therefore our derived
DLA bias factor implies a condition on the characteristic mass of halos
hosting DLAs. We use the model of \citet{Tinker+2010} to
calculate the halo bias at the mean redshift of our cross-correlation
measurement $z=2.3$, shown as the thick solid curve in figure
\ref{fig DLA: dla halo mass} (both left and right panels). The grey
horizontal line with the shaded band is the value of $b_{\rm DLA}$ for
our C1 sample and $r_{\rm min} = 5\hmpc$, with the bootstrap error. This
is the result with the smallest error that we believe we can trust, as
we have argued above. However, it does not include the systematic error
arising mainly from the \lya\ bias factors and impurities in the catalog.

 If all DLAs were in halos with a single value of the mass, the
inferred halo mass would be placed from $2.5\cdot 10^{11}\hmsun$ to
$5\cdot 10^{11}\hmsun$ given our statistical error bar, corresponding
to a massive galaxy. However, in any realistic model, DLAs
host halos should have a broad mass range. Our measurement yields only
the mean bias factor, which depends on the DLA cross section as a
function of halo mass as discussed in FR12. This cross section depends
on the distribution of gas in halos, and therefore on the complex
physics of gas accretion, galaxy formation, and galactic and quasar
winds that can expel gas from a central galaxy to the outer regions of
halos or to the intergalactic medium.

  Following FR12, we assume a power-law distribution of the DLA
cross-section $\Sigma(M_h)$ as a function of halo mass,
\begin{equation}
	\label{eq DLA: DLA cross-section} 
	\Sigma(M_h) = \Sigma_{0}\left(\frac{M_h}{M_{\rm min}}\right)^{\alpha}
\,\,\,\,\left(M>M_{min}\right)~,
\end{equation}
The predicted mean DLA bias under this simple assumption is
\begin{equation}
 b_{\rm DLA} = { \int_{M_{\rm min}}^{M_{\rm max}}
  n(M_h) \Sigma(M_h) b(M_h)\, dM_h
 \over \int_{M_{\rm min}}^{M_{\rm max}} n(M_h) \Sigma(M_h) \, dM_h } ~,
\end{equation}
where $n(M_h)$ is the number density and $b(M_h)$ the bias of
halos of mass $M_h$. We have also assumed that the cross section is
negligible below a minimum mass $M_{\rm min}$ and above a maximum mass
$M_{\rm max}$. Numerical simulations of galaxy formation including
hydrodynamics and complex recipes for star formation and galaxy winds
driven by supernova explosions have been extensively studied by several
groups \citep[e.g.,][]{Pontzen+2008,Tescari+2009,Bird+2014}, and can predict this
relationship. In particular, \cite{Bird+2014} find a relation that is
well fitted by a power-law over the halo mass range that can be probed
by their simulations, $10^{8.5}\hmsun < M < 10^{12}\hmsun$. At lower
masses, the intergalactic photoionized gas has sufficient pressure to
slow the accretion onto halos. At higher halo masses, the small box
of their simulations do not allow enough halos to be included to derive
solid predictions.

\begin{figure*}
	\centering
	\includegraphics[width=\textwidth]{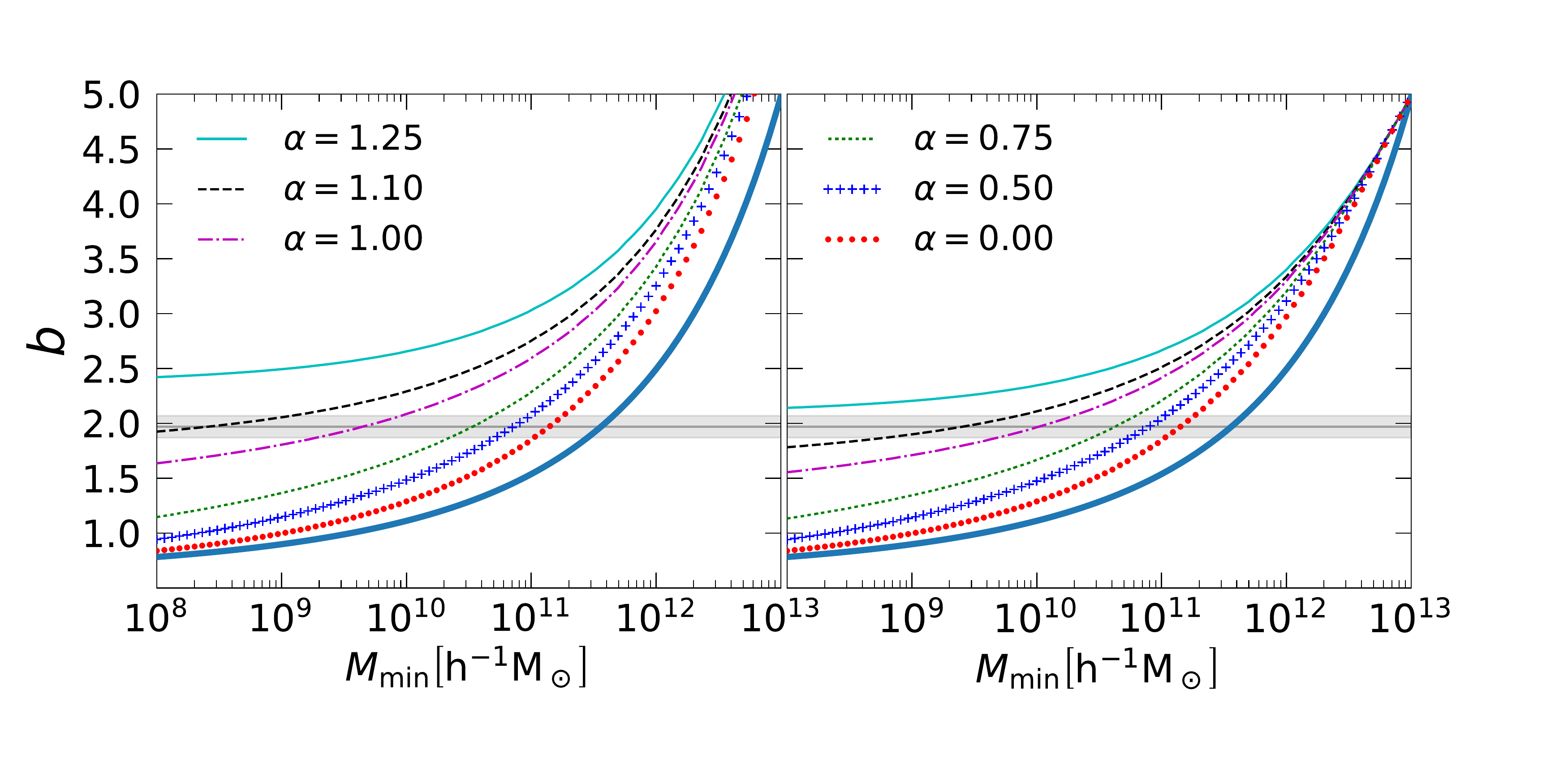}
	\caption{{\it Left:} Average DLA bias when the
DLA cross-section as a function of halo mass
follows the power-law relation in equation (\ref{eq DLA: DLA cross-section}),
for the indicated values of the power-law index $\alpha$, as a function
of the lower mass cutoff $M_{\rm min}$.
{\it Right:} Same as in the left panel but including an upper mass
cutoff at $M_{max}=10^{13}\hmsun$. The bias for a single halo mass is shown
in both panels as the thick solid line. All cases are computed at
$z = 2.3$. Horizontal shaded region is our derived value for the
DLA bias (sample C1, $r_{\rm min}=5\hmpc$) and the 1$\sigma$
statistical bootstrap error (not including errors on the \lya{} forest
bias factors).}
	\label{fig DLA: dla halo mass}
\end{figure*}
 
  The mean DLA bias computed in this model is shown in the left panel of
figure \ref{fig DLA: dla halo mass} as a function of $M_{\rm min}$, and
in the limit of infinite $M_{\rm max}$, for six different values of the
power-law index $\alpha$. For $M_{min}=10^{8.5}\hmsun$, the required
slope to match our DLA bias is $\alpha\sim 1.1$. The simulations of
\cite{Bird+2014} tend to give a lower slope of $\alpha\simeq 1.0$, which
corresponds to $b_{\rm DLA}\simeq 1.7$ for the same value of
$M_{\rm min}$, which is discrepant with our measurement at the
$\sim 2-\sigma$ level if we use only our statistical, bootstrap errors,
but may be more consistent with what we measure when including plausible
systematic errors from uncertainties in the \lya\ forest bias factors
and the effects of catalog impurity. If a bias factor $b_{\rm DLA} >
1.9$ is confirmed by future by improved determinations, then either a
steeper slope or higher value of $M_{\rm min}$ compared to the
\cite{Bird+2014} simulations would be required. Results of earlier
simulations that had weaker galactic winds and predicted slopes of
$\alpha\simeq 0.7$ \citep{Pontzen+2008}, with much lower implied DLA
bias factors of $b_{\rm DLA} < 1.5$, are strongly ruled out by our
measurement. In general, the high value of $b_{\rm DLA}$ we measure
implies that the structure of DLAs is affected by strong galactic
winds, which are able to decrease the cross sections in low-mass halos
by expelling gas to the intergalactic medium, and increase cross
sections in high-mass halos by spreading gas out to large radius.

 However, \cite{Bird+2014} assumed an extrapolation of their power-law
fit to $M_h > 10^{12} \hmsun$ to derive a bias factor
$b_{\rm DLA}\simeq 1.7$ from their fitted power-law slope, because their
small simulations cannot predict the properties of the rare, more
massive halos. These massive halos are very highly biased and make an
important contribution to the mean bias of DLAs. The right panel of
figure \ref{fig DLA: dla halo mass} shows the same models using now an
upper cutoff $M_{\max}=10^{13} \hmsun$, and we see that in this case,
the predicted bias factor for $M_{\rm min}=10^{8.5} \hmsun$ and
$\alpha=1$ already decreases to $b_{\rm DLA}\simeq 1.6$. The results of
\cite{Bird+2014} can therefore agree with our measurement at better than
$2-\sigma$ only if we allow for a systematic error and if there is no
substantial flattening of the slope of the $\Sigma(M_h)$ relation at
$M_h > 10^{12} \hmsun$. There are reasons to expect this flattening of
the slope because at low redshift, we observe that massive halos are
associated with galaxy groups and clusters containing most of the
baryons in X-ray emitting hot gas, where much of the cold gas in
galaxies is destroyed owing to tidal and ram-pressure stripping
\citep{Fabian2012}. However, at the redshifts where DLAs in
BOSS are found, the amount of cold gas in very massive halos is
not well known. Larger and better simulations, and observations of
galaxy clusters at high redshift, are required to clarify this question.

 Our improved measurement of the bias factor of DLAs has an
impact on forecasts for 21-cm surveys of HI galaxies
\citep[see, e.g.,][]{Chang+2010,Chang+2015,Castorina+2016,Villaescusa+2016}: a higher
value of the bias implies a larger amplitude of the 21-cm
fluctuations. Our new value is very similar to the previous one by FR12,
with a reduced error and a more detailed analysis of model dependences.
The 21-cm fluctuation amplitude depends on the mean bias of all
absorbers weighted by their HI column density. The lack of dependence of
$b_{\rm DLA}$ on $N_{\rm HI}$, and the fact that most of the known
neutral hydrogen in the Universe resides in DLAs, strongly suggest that
our derived value $b_{\rm DLA}\simeq 2.0$ should apply for the neutral
gas that will be detected in 21-cm surveys. Although these surveys
should include dust-absorbed systems that are not included in our DLA
sample and lower column density systems that we also do not include, it
is difficult that these systems may change the mean bias factor
appreciably.

 Finally, we comment on one theoretical aspect of the bias of dark matter
halos that may influence the comparison of the theoretically predicted
and observed DLA bias factor. The halo bias is not only a function of
mass, but also of the assembly history of a halo, a phenomenon known as
``assembly bias'' \citep[e.g.,][]{Borzyszkowski+2017}. Halos of a fixed
mass in high-density regions tend to have accreted their mass recently,
whereas in low-density regions the accretion rate is lower. The DLA
cross section may depend also on the accretion history: for a fixed halo
mass, a high accretion rate may imply more atomic gas is available at
large radius to give rise to a DLA system, and at the same time a higher
bias owing to the assembly bias effect. This effect may be missed by
simulation results like those in \cite{Bird+2014} when the bias factor
is inferred from the DLA host halo mass distribution and the same type
of theoretical relation of bias and halo mass we have used here, instead
of being directly obtained from the simulation. 
Alternatively one may achieve a steeper slope cross-correlation - halo mas relation, and then get a higher predicted DLA bias, by changing some parameters of the winds models in simulation.
Future studies should
therefore also attempt to include the effect of assembly bias or wind model when
comparing to the observational result.
%
\section{Summary and Conclusions}  \label{sec DLA: Conclusions}

 We have measured the cross-correlations of DLAs and the \lya{}
forest for several samples of DLAs of the final DR12 of
BOSS: 23,342 DLAs with $N_{HI} \ge 10^{20}\cm^{-2}$ in
the redshift range $2.0\ge z_{\rm DLA}\ge3.5$. We have found that the
simple linear theory model for this cross-correlation, with the redshift
distortions predicted by \cite{Kaiser1987}, is fully consistent with the
data, and we have obtained the DLAs bias factor required to match
the measured cross-correlation amplitude. Our main conclusions are as
follows:

\begin{itemize}
	\item We measure $b_{\rm DLA} = (1.99\pm 0.11)$ for sample C2,
extending the fit range down to $r_{\rm min}=5\hmpc$. A more
conservative result, using sample A and $r_{\rm min}=10\hmpc$ to avoid
possible non-linear effects, yields $b_{\rm DLA} = (2.00\pm 0.19)$. Both
values are similar to the previous result reported by FR12, but the
detailed comparison depends on several differences in the analysis and
model dependences discussed in section \ref{sec DLA: comparison DR9}.
	
	\item We do not find any dependence of the DLAs bias on redshift
and $N_{\rm HI}$, at the level of $\sim 10\%$ over the ranges $2<z<3$ and
$20 < \log N_{\rm HI} < 21.5$. The independence on redshift suggests
that the characteristic host halo mass is decreasing with redshift, and
the independence with column density suggests that the mean $N_{\rm HI}$
radial profile is similar in host halos of different masses.

	\item The value of the DLAs bias does not significantly
change among our samples that include or exclude DLAs in the \lyb{}
forest or near the \lya{} and \ovi{} quasar emission lines, suggesting that
systematics associated with these cuts are small.

	\item We detect no scale dependence in the DLAs bias,
which reinforces the agreement of the measured cross-correlation with
the linear model we assume, based on the $\Lambda CDM$ power spectrum
with the parameters determined by \cite{Planck2015}.

	\item The principal systematic errors that need to be addressed
to make the measurement of $b_{\rm DLA}$ more robust are the dependence
on the \lya{} forest bias parameters and the HCD correction, and the
effects of impurities and selection effects in the DLAs catalogue. The
absence of any dependence on column density, and the small variations
of the DLA bias with the HCD correction when used consistently in the
same models that fit the \lya{} autocorrelation results of
\cite{Bautista+2017} suggests that these systematics are not larger than
our statistical errors. The small variation of the high-ionization lines
mean equivalent width with $N_{\rm HI}$ found in \cite{Mas-Ribas+2017}
also suggest the same thing.

	\item Assuming the DLA cross section versus halo mass relation
$\Sigma(M_h) \propto M_h^{\alpha}$ down to
$M_{\rm min} \sim 10^{8.5} \hmsun$, we find that $\alpha > 1$ is
required to match the observed $b_{\rm DLA}$, a steeper relation than is
predicted in most numerical simulations of galaxy formation. Even for
the simulations with strong winds analysed by \cite{Bird+2014},
which predict a steeper relation than previous models, the implied bias
is only marginally consistent with our observational determination, and
needs to assume an extrapolation of this power-law relation with
$\alpha\simeq 1$ up to halo masses much larger than the ones being
probed by their simulation results. The effect of assembly bias may
increase the theoretical prediction for $b_{\rm DLA}$ and help bringing
it into agreement with our observational determination. 		
\end{itemize}

\vspace{6mm}

\section*{Acknowledgments}
Funding for SDSS-III has been provided by the Alfred P. Sloan Foundation, the Participating Institutions, the National Science Foundation, and the U.S. Department of Energy Office of Science. The SDSS-III web site is http://www.sdss3.org/. 

SDSS-III is managed by the Astrophysical Research Consortium for the Participating Institutions of the SDSS-III Collaboration including the University of Arizona, the Brazilian Participation Group, Brookhaven National Laboratory, University of Cambridge, Carnegie Mellon University, University of Florida, the French Participation Group, the German Participation Group, Harvard University, the Instituto de Astrofisica de Canarias, the Michigan State/Notre Dame/JINA Participation Group, Johns Hopkins University, Lawrence Berkeley National Laboratory, Max Planck Institute for Astrophysics, Max Planck Institute for Extraterrestrial Physics, New Mexico State University, New York University, Ohio State University, Pennsylvania State University, University of Portsmouth, Princeton University, the Spanish Participation Group, University of Tokyo, University of Utah, Vanderbilt University, University of Virginia, University of Washington, and Yale University.

IPR and JME were supported by the Spanish MINECO under
projects AYA2012-33938 and AYA2015-71091-P and MDM-2014-0369 of ICCUB (Unidad de Excelencia 'María de
Maeztu'). AFR is supported by a STFC Rutherford Fellowship, grant reference ST/N003853/1. SB was supported by NASA through Einstein Postdoctoral Fellowship Award Number PF5-160133.

\bibliographystyle{apj}
\bibliography{iprafols}

\appendix
%
\section{Public access to data and code}\label{sec DLA: public access}

 The software used to generate the results in this paper is written in
$C^{++}$ and is publicly available at
{\url{https://github.com/iprafols/cross_correlations}. This repository
also contains a Python library with functions to plot the
cross-correlation measurements, and the correlation function and
configuration files necessary to reproduce our main results.
Instructions to install and run the software are in the repository.
Data and configuration files are in plain text format.

%
\section{Projector of the \texorpdfstring{$\delta$}{delta} field and the distortion matrix formalism}\label{sec DLA: proj}

\subsection{Motivation}\label{subs DLA: motivation}

As explained in section \ref{sec DLA: Cross-correlation}, the assumption is that the measured \lya{} transmission fluctuation, $\delta^{(m)}$, differs from the true \lya{} transmission fluctuation, $\delta^{(t)}$ according to 
\begin{equation}
	\label{eq DLA: hypothesis}
	\delta_{i}^{(m)} = \delta_{i}^{(t)}+a+b\lambda_{i} ~,
\end{equation}
where $a$ and $b$ are small unknown functions that depend on the $\delta$ field in a complicated manner, and $\lambda$ is either the wavelength or the logarithm of the wavelength (whichever is used in the computation of the $\delta$ field). Here we assume that $a$, $b$ are constant within a given forest.

This hypothesis is motivated by the definition of the $\delta$ field. As explained in section \ref{subs DLA: lya Sample} the $\delta$ field is defined as
\begin{equation}
	\label{eq DLA: delta}
	\delta_{i} = \frac{f_{i}}{C_{q}\left(\lambda_{i}\right)\overline{F}\left(z_{i}\right)} - 1 ~,
\end{equation}
where $f_{i}$ is the measured flux, $C_{q}\left(\lambda_{i}\right)$ is the quasar continuum (or unabsorbed flux), and $\overline{F}\left(z_{i}\right)$ is the mean transmitted fraction at the \lya{} absorber redshift. The pixel redshift is $z_{i}=\lambda_{i}/\lambda_{\rm Ly\alpha} - 1$. The quasar continuum is assumed to have the form $C_{q}\left(\lambda_{i}\right)=\overline{C}\left(\lambda_{i}\right)\left(a+b\lambda_{i}\right)$, where $\overline{C}$ is the mean flux determined by stacking all quasar spectra, estimated at the restframe wavelength, and $a$ and $b$ are fitted constants, different for different forests. We can fit the parameters $a$ and $b$ except for a small error, i.e., $a = a_{t}-\delta_{a}$ and $b = b_{t}-\delta_{b}$.

If we Taylor expand this expression and retain only the leading order
\begin{equation}
	\delta^{(m)}_{i} \approx \delta^{(t)}_{i} - \frac{\delta_{a}}{a_{t}+b_{t}\lambda_{i}} - \frac{\delta_{b}\lambda}{a_{t}+b_{t}\lambda_{i}} ~.
\end{equation}
We can now assume that the average of $b_{t}$ along the different forests will be zero and that for each individual forest it is a small fluctuation of this average. This assumption is motivated by the fact that the steepness of the flux spectra is accounted for in the estimation of $\overline{C}$. Therefore, we can neglect $b_{t}\lambda$ over $a$, hence the presented hypothesis (equation \ref{eq DLA: hypothesis}).

\subsection{Projector}\label{subs DLA: derivation}
Since it is impossible to know the values of $a$ and $b$ in equation \ref{eq DLA: hypothesis}, it is necessary to indentify a projector, $P$, that allows the removal of these parameters, i.e.,
\begin{equation}
	\label{eq DLA: condition 2}
	P\delta^{(m)} = P\delta^{(t)} ~.
\end{equation}
To find an expression for this projector, it is useful to adopt a vectorial representation, which allows one to treat the forest as a whole. Keep in mind that we are assuming $a$ and $b$ to be constant throughout the forest.

To start with the derivation we first consider the case $b=0$. In vectorial form, and for a forest of length $N$, we have
\begin{equation}
	\label{eq DLA: hypothesis2}
	\left(\begin{matrix} \delta^{(m)}_{1} \\ \vdots \\ \delta^{(m)}_{N}\end{matrix}\right) = \left(\begin{matrix} \delta^{(t)}_{1} \\ \vdots \\ \delta^{(t)}_{N}\end{matrix}\right) - av_{1}~,
\end{equation}
where $v_{1}$ is the vector $\frac{1}{\mathcal{N}_{1}}\left(1, \cdots, 1\right)$, and $\mathcal{N}_{1}$ is a normalization constant that makes the vector unitary, i.e. $v_{1}^{t}v_{1} = 1$.

We now construct a projector $P_{0}$ which will cancel the second term in the equation above. This projector reads
\begin{equation}
	\label{eq DLA: projector0}
	P_{0} = \mathbb{I}-v_{1}v_{1}^{t} ~,
\end{equation} 
and it does indeed cancel the second term in equation\ref{eq DLA: hypothesis2}:
\begin{equation}
	\label{eq DLA: test P0}
	P_{0}v_{1} = \left(\mathbb{I}-v_{1}v_{1}^{t}\right)v_{1} = \mathbb{I}v_{1}-v_{1}\underbrace{v_{1}^{t}v_{1}}_{=1} = v_{1}-v_{1} = 0 ~.
\end{equation}

Now that we have an appropriate projector, let us relax the condition $b=0$. We now have
\begin{equation}
	\left(\begin{matrix} \delta^{(m)}_{1} \\ \vdots \\ \delta^{(m)}_{N}\end{matrix}\right) = \left(\begin{matrix} \delta^{(t)}_{1} \\ \vdots \\ \delta^{(t)}_{N}\end{matrix}\right) - av_{1} - bv_{2}~,
\end{equation}
where $v_{2} = \left(\lambda_{1}, \cdots, \lambda_{N}\right)$. 

We must expand the projector $P_{0}$ to a new projector $P$ in such a manner that maintains the condition $Pv_{1}=0$ imposed in the particular case where $b=0$ and add the extra condition that $Pv_{2} = 0$, i.e., we have to project using a vector which is orthogonal to $v_{1}$. We can follow the Gram-Schmidt process to determine such a vector: $u_{2}=v_{2}-\left(v_{2}^{t}v_{1}\right)v_{1}$. Any vector in this direction will verify $P_{1}v_{1}=0$. However, for it to verify $Pv_{2}=0$, we need a vector in the direction of $u_{2}$ which is properly normalized. Therefore the new projector reads
\begin{equation}
	\label{eq DLA: projector1}
	P = \mathbb{I}-v_{1}v_{1}^{t}-\frac{1}{\mathcal{N}_{2}^{2}}u_{2}u_{2}^{t} ~,
\end{equation}
where $\mathcal{N}_{2}^{2} = u_{2}^{t}u_{2}= v_{2}^{t}v_{2}-v_{2}^{t}v_{1}v_{1}^{t}v_{2}$.

This projector verifies both conditions:
\begin{multline}
	Pv_{1} = \left[\mathbb{I}-v_{1}v_{1}^{t}-\frac{1}{\mathcal{N}_{2}^{2}}u_{2}u_{2}^{t}\right]v_{1} = 
\\
	=\underbrace{P_{0}v_{1}}_{=0} - \frac{1}{\mathcal{N}_{2}^{2}}u_{2}u_{2}^{t}v_{1} = -\frac{1}{\mathcal{N}_{2}^{2}}\left[v_{2}-\left(v_{2}^{t}v_{1}\right)v_{1}\right] \times
\\
	\times\left[v_{2}-\left(v_{2}^{t}v_{1}\right)v_{1}\right]^{t}v_{1} =-\frac{1}{\mathcal{N}_{2}^{2}}\left[v_{2}-\left(v_{2}^{t}v_{1}\right)v_{1}\right]\times
\\
	 \times\underbrace{\left(v_{2}^{t}v_{1}-\left(v_{2}^{t}v_{1}\right)\underbrace{v_{1}^{t}v_{1}}_{=1}\right)}_{=0} = 0 ~,
\end{multline}
and
\begin{multline}
	Pv_{2} = \left[\mathbb{I}-v_{1}v_{1}^{t}-\frac{1}{\mathcal{N}_{2}^{2}}u_{2}u_{2}^{t}\right]v_{2} = \mathbb{I}v_{2}-v_{1}v_{1}^{t}v_{2}- 
\\
	-\frac{1}{\mathcal{N}_{2}^{2}}\left[v_{2}-\left(v_{2}^{t}v_{1}\right)v_{1}\right] \left[v_{2}-\left(v_{2}^{t}v_{1}\right)v_{1}\right]^{t}v_{2}=
\\
	= v_{2}-v_{1}v_{1}^{t}v_{2}-\frac{1}{\mathcal{N}_{2}^{2}}\left[v_{2}-\left(v_{2}^{t}v_{1}\right)v_{1}\right]\times
\\
	\times\underbrace{\left(v_{2}^{t}v_{2}-\left(v_{2}^{t}v_{1}\right)v_{1}^{t}v_{2}\right)}_{=\mathcal{N}^{2}} = =v_{2}-v_{1}\left(v_{1}^{t}v_{2}\right)-
\\
	-\left(v_{2}-v_{1}\left(v_{1}^{t}v_{2}\right)\right) = 0 ~.
\end{multline}

This projector allows us to compare the real and the measured values without knowing the parameters $a$ and $b$. The derivation has been performed without specifying the scalar product, so the expression for the projector (equation \ref{eq DLA: projector1}) is then valid for any given scalar product. This behaviour is interesting because not all the pixels in the \lya{} forest are equally noisy; there is a weight associated with each pixel. This weight is now easily introduced into this formalism if one simply defines the scalar product as
\begin{equation}
	\label{eq DLA: scalar product}
	u^{t}v = \sum_{i\in f}u_{i}v_{i}w_{i}~,
\end{equation}
where $i$ is an index that runs over pixels in a particular forest $f$.

Now that we have specified the scalar product, we can find specific expressions for $v_{1}$, $u_{2}$, and $\mathcal{N}_{2}^{2}$.
\begin{multline}
	\left(\begin{matrix} 1 & \cdots & 1\end{matrix}\right) \left(\begin{matrix} 1 \\ \vdots \\ 1\end{matrix}\right)= \sum_{i\in f}w_{i} \Rightarrow
\\
	\Rightarrow v_{1} = \frac{1}{\sqrt{\sum_{i\in f} w_{i}}}\left(1, ..., 1\right) ~,
\end{multline}
\begin{multline}
	u_{2} = v_{2}-v_{2}^{t}v_{1}v_{1} = v_{2}-\frac{\sum_{i\in f} \lambda_{i}w_{i}}{\sqrt{\sum_{i\in f} w_{i}}}v_{1} = 
\\	
	=v_{2} - \bar{\lambda}\left(\begin{matrix} 1 \\ \vdots \\ 1\end{matrix}\right) = \left(\begin{matrix} \lambda_{1}-\bar{\lambda} \\ \vdots \\ \lambda_{N}-\bar{\lambda}\end{matrix}\right) ~, \text{ and}
\end{multline}
\begin{multline}
\mathcal{N}_{2}^{2} = u_{2}^{t}u_{2} = \sum_{i\in f}\left(\lambda_{i}-\bar{\lambda}\right)^{2}w_{i} ~,
\end{multline}
where $\bar{x}\equiv\sum_{i\in f}xw_{i}/\sum_{i\in f}w_{i}$.

Using this scalar product and the corresponding expressions for $v_{1}$, $u_{2}$, and $\mathcal{N}_{2}^{2}$ derived above, we can study the behaviour of this projector when it is applied to a vector $\delta$, defined in the forest of interest.
\begin{multline}
	P\delta = \left[\mathbb{I}-v_{1}v_{1}^{t}-\frac{1}{\mathcal{N}^{2}}u_{2}u_{2}^{t}\right]\delta = \delta - v_{1}v_{1}^{t}\delta-
\\
 	-\frac{1}{\mathcal{N}_{2}^{2}}u_{2}u_{2}^{t}\delta = \delta - \frac{\sum_{i\in f}\delta_{i}w_{i}}{\sum_{i\in f} w_{i}}\left(\begin{matrix} 1 \\ \vdots \\ 1\end{matrix}\right)- 
\\
 	-\frac{\sum_{i\in f}\left(\lambda_{i}-\bar{\lambda}\right)\delta_{i}w_{i}}{\sum_{i\in f}\left(\lambda_{i}-\bar{\lambda}\right)^{2}w_{i}} 	\left(\begin{matrix} \lambda_{1}-\bar{\lambda} \\ \vdots \\ \lambda_{N}-\bar{\lambda}\end{matrix}\right) ~.
\end{multline}
As we will see later, it is useful to consider the $i^{\rm th}$ component of this vector:
\begin{multline}
	\left(P\delta\right)_{i} = \sum_{j\in f}P_{ij}\delta_{j} = \delta_{i} -\bar{\delta} - \frac{\sum_{j\in f}\delta_{j} \left(\lambda_{j}-\bar{\lambda}\right)w_{j}}{\sum_{j\in f}\,\left(\lambda_{j}-\bar{\lambda}\right)^{2}w_{j}} \times
\\ 
	\times\left(\lambda_{i}-\bar{\lambda}\right)~.
\end{multline}

\subsection{Distortion matrix}\label{subs DLA: distortion matrix appendix}
The $\chi^{2}$ statistic for this estimator reads
\begin{equation}
	\label{eq DLA: chi2}
	\chi^{2} = \left(\xi-\left<\xi\right>\right)^{t}C^{-1}\left(\xi-\left<\xi\right>\right) ~,
\end{equation}
where $C$ is the covariance matrix between the different bins of the cross-correlation. 

The expected value of the cross-correlation estimator in bin $A$ can be written as
\begin{equation}
	\label{eq DLA: <xi>}
	\left<\xi^{A}\right> = \frac{\sum_{d,f}\sum_{i\in f}\Theta^{A}_{id}w_{i}\sum_{j\in{f}} P_{ij}\xi_{jd}}{\sum_{d,f}\sum_{i\in f}\Theta^{A}_{id}w_{i}} ~,
\end{equation}
where the indexes $d$ and $f$ run over DLAs and forests, respectively, the indexes $i$ and $j$ run over pixels in a particular forest, $\Theta_{id}^{A}$ is 1 if the DLA-pixel pair is in bin $A$ and 0 otherwise, and $\xi_{jd}$ is the theoretical prediction of the cross-correlation for the DLA-pixel pair $jd$.

At this point we can discretize the model similarly to the discretization of the data. Then 
\begin{equation}
	\xi_{jd} = \sum_{B}\Theta_{jd}^{B}\xi^{B} ~.
\end{equation}
This discretization need not be the same as the discretization on the data, but it is convenient to do so. The formalism presented here applies to whichever case is chosen.

Introducing this discretization into equation \ref{eq DLA: <xi>}, the expected value of the cross-correlation can be written as
\begin{multline}
	\left<\xi^{A}\right> = \sum_{B}\frac{\sum_{d,f}\sum_{i\in f}\Theta^{A}_{id}w_{i}\sum_{j\in{f}} P_{ij}\Theta^{B}_{jd}\xi^{B}}{\sum_{d,f}\sum_{i\in f}\Theta^{A}_{id}w_{i}} \equiv 
\\
	\equiv\sum_{B}D^{AB}\xi^{B} ~, 
\end{multline}
where $D^{AB}$ is the distortion matrix element that relates the cross-correlation measured in bin $A$ and the model for the cross-correlation in bin $B$. The quantity $D^{AB}$ is defined as

\begin{multline}
	D^{AB} = \frac{1}{\sum_{d,f} \sum_{i\in f}\Theta_{id}^{A}w_{i}}\sum_{d,f} \sum_{i\in f}\Theta_{id}^{A}w_{i} \sum_{j\in f} \left(\delta_{ij} - \right.
\\
	 -\left.\frac{w_{j}}{\sum_{k\in f}w_{k}}-\frac{\left(\lambda_{j}-\bar{\lambda}_{f}\right)\left(\lambda_{i}-\bar{\lambda}_{f}\right)w_{j}}{\sum_{k\in f}\left(\lambda_{k}-\bar{\lambda}_{f}\right)^{2}w_{k}}\right)\Theta_{jd}^{B}
	 ~,
\end{multline}
where $\delta_{ij}$ is the Kronecker delta and should not be confused with the \lya{} transmission fluctuation.

\end{document}